\documentclass{aa}
\usepackage[varg]{txfonts}
\usepackage{natbib}
\bibpunct{(}{)}{;}{a}{}{,} % to follow the A&A style
\usepackage[colorlinks=true,     linkcolor=blue, citecolor=blue, filecolor=blue, urlcolor=blue]{hyperref}
\usepackage{graphicx} % Required for inserting images
\usepackage{amsmath}
\usepackage[T1]{fontenc}
\usepackage{rotating}
\usepackage[table,xcdraw,pdftex,dvipsnames]{xcolor}
\usepackage{adjustbox,placeins}
\usepackage{stfloats}

\newcommand{\rev}[1]{#1} % for version without revisions highlighted
\newcommand{\revtwo}[1]{#1} % for version without revisions highlighted
\newcommand{\reveditor}[1]{#1} % for version without revisions highlighted
\begin{document}
\nolinenumbers

\title{WR + O binaries as probes of the first phase of mass transfer}

\author{Marit Nuijten
          \inst{1}
          \and
          Gijs Nelemans\inst{1, 2, 3}
          }

\authorrunning{M. Nuijten \& G. Nelemans}

\institute{
             Department of Astrophysics/IMAPP, Radboud University, PO Box 9010, 6500 GL, The Netherlands
        \and
Institute of Astronomy, KU Leuven, Celestijnenlaan 200D, 3001 Leuven, Belgium
         \and           
            SRON, Netherlands Institute for Space Research, Niels Bohrweg 4, 2333 CA Leiden, The Netherlands
             }

\abstract
{}{Wolf-Rayet (WR) and O-star binaries can be the progenitors of X-ray binaries and double black hole binaries. Their formation is not yet fully understood, however. For \rev{21} observed WR+O systems, we aim to infer \rev{whether the mass transfer started on the main sequence (\revtwo{C}ase A) or later (\revtwo{C}ase B).  We also calculated (limits on) the mass-transfer efficiency $\beta$, that is, the fraction of transferred mass that is accreted, and the parameter $\gamma$, which denotes the fraction of angular momentum \revtwo{of the binary} that is lost per unit mass in units of the average angular momentum \revtwo{of the binary} per unit mass.}}
{We inferred the possible values for the initial masses based on the observed WR masses and models for WR from the literature. With these initial primary masses, we created a grid of possible periods and secondary masses for which we determined the values that $\beta$ and $\gamma$ would have taken for either \revtwo{C}ase A or \revtwo{C}ase B mass transfer. Based on this, we also determined the case of mass transfer that is most likely for each system.}{Taking into account the progenitor distribution of WR+O binaries, we find that highly non-conservative \rev{\revtwo{C}ase A} mass transfer seems to be the most likely scenario \rev{for the majority of systems as this can explain 14 out of 21 systems}. The angular momentum \rev{loss} is likely relatively high (typically \rev{$\gamma > 1$}).  \rev{Our finding that most systems in our sample experienced Case A mass transfer contradicts the expectation that most massive binaries go through \revtwo{C}ase B mass transfer. This} suggests that post-case-B systems are significantly underrepresented in the observed WR+O binary population\rev{, either intrinsically or due to severe selection effects}.}{}

\maketitle
\nolinenumbers

\section{Introduction}
Almost all massive stars ($\gtrsim 95\%$) have one companion or more than one \citep[see e.g.][]{2011A&A...530A.108E, 2013A&A...550A.107S, 2013A&A...550A..27M, 2014ApJS..213...34K,Moe2017,2017A&A...598A..84A, 2017IAUS..329...89B, 2021MNRAS.507.5348V, 2021A&A...655A...4T, 2022A&A...658A..69B,2024BSRSL..93..170M}. Roughly two-thirds of the massive stars in a binary are close enough to their companion that they will interact during their lifetime \citep{Sana_2012}. This leads to many interesting phenomena. It is for instance possible for both massive stars in a binary to become a black hole \citep[see for a review][]{2022LRR....25....1M}. A binary with two black holes is an interesting object to study in relation to, for example, the large number of gravitational wave detections of double black holes \citep{2023PhRvX..13d1039A}.  

The formation path for black hole binaries and also high-mass X-ray binaries \citep[e.g.][]{Tauris2003} typically involves a first mass-transfer phase that can have crucial consequences for the following evolution \citep[see][]{2024MNRAS.530.3706D}. In this paper, we focus on binaries that contain a Wolf-Rayet (WR) star and an O-star that likely have experienced this phase \citep[e.g.][]{1967AcA....17..355P,1998NewA....3..443V}.  \rev{WR stars are typically viewed as stars} that lost (most of) their envelope and therefore contain little to no hydrogen on their surface, but have a high abundance of other elements\rev{, although some were found to be main-sequence stars \citep[see e.g.][]{2023A&A...680A..22M}}. In general, the mass of WR stars ranges from 10 to 25 $M_{\odot}$, and an O-star is their progenitor \citep{Crowther2006}. WR+O binaries can be the progenitors of interesting phenomena such as X-ray binaries and the aforementioned black hole-black hole binaries \citep[e.g.][]{Tauris2003,Van_den_Heuvel2017}. However, in order to understand the formation of WR+O binaries and their potential to evolve into X-ray binaries and double black holes, the mass-transfer process must be understood. Two of the key ingredients of this process are the fraction of mass that is accreted by the accreting star and the fraction that is lost from the system (the mass-transfer efficiency), and the amount of angular momentum that this lost material that was removed from the system. The efficiency of the mass transfer can strongly affect the evolution of the stars within the binary and the binary as a whole (see e.g. \cite{Shao2016}). We studied whether it is possible to determine the values of the mass-transfer efficiency and angular momentum loss of these systems and \rev{the evolutionary phase of the donor at the start of mass transfer}. Because of the significant mass loss, WR+O stars can in principle also form because the WR progenitor loses its envelope via a stellar wind. However,  the properties of the observed systems, such as the spins of the O stars and the orbital properties, suggest that the majority of WR+O binaries have experienced mass transfer \citep{2017MNRAS.464.2066S, 2018ARep...62..567C,2018A&A...615A..65V}.

Mass transfer is classified depending on the phase in the evolution of the donor the mass in which transfer starts. The first type of mass transfer is \revtwo{C}ase A. This is when the donor star is still on the main sequence (MS) at the moment it fills its Roche lobe \citep[e.g.][]{1994A&A...290..119P,2001A&A...369..939W, 2001ApJ...552..664N, 2007A&A...467.1181D, 2010A&A...510A..13V, 2022A&A...659A..98S}. The phase when hydrogen burning in the core of the star has stopped but helium burning has not yet started at the onset of mass transfer is called \revtwo{C}ase B mass transfer \citep[e.g.][]{1967AcA....17..355P, 2011A&A...528A..16V, 1992ApJ...391..246P, 2022A&A...662A..56K}. There are also cases when \revtwo{C}ase A is followed by a second phase that is \revtwo{C}ase B. These cases are called \revtwo{C}ase AB mass transfer \citep[see e.g.][]{2022A&A...659A..98S}. When mass transfer starts after helium ignition in the core, it is called  Case C \citep[e.g.][]{1992ApJ...391..246P}. We focused on the first two cases of mass transfer because Case C mass transfer is expected to be unstable more often than \revtwo{C}ase A or \revtwo{C}ase B \citep[e.g.][]{Ge2015}, and we studied the formation of WR+O binaries via stable mass transfer.

\cite{Petrovic2005} studied three WR+O binaries and concluded for all three that they must have undergone \revtwo{C}ase A mass transfer. \cite{Shao2016} also studied different WR+O systems based on models, but were unable to draw a clear conclusion on which type of system undergoes \revtwo{C}ase A or \revtwo{C}ase B mass transfer. \citet{2022A&A...659A..98S} studied the first mass-transfer phase in detail and compared their models to observations of systems that are currently transferring mass, but also showed results for the systems after the mass transfer. \rev{However, all these studies compared the observed systems to one or more specific binary models. We take a different approach and derive for each observed WR+O binary the type of evolution could have led to the current system. In this way, we constrained the fraction of \revtwo{the transferred mass} that was lost from the system and the specific angular moment it took with it.}

We used the VIIth catalogue of Galactic WR stars \citep{vanderHucht2001} to \rev{choose} suitable candidates for our study \rev{(Sect.~\ref{Objects})}. This resulted in a sample of \rev{21} WR+O binaries and allowed us to take a more statistical approach to study their mass transfer. This paper is structured as follows. In Sect.~\ref{Method} we introduce the method with which we constrained the possible progenitors of the observed WR+O binaries and the resulting mass-transfer efficiency, angular momentum loss, and mass-transfer type. In Sect.~\ref{Objects} we introduce the properties of the binaries we studied. In Sect.~\ref{Results} we present the results, which we discuss in Sect.~\ref{discussion} and summarise in Sect.~\ref{conclusion}

\section{Method}\label{Method}

We inferred the mass-transfer type and (limits on) the mass-transfer efficiency and angular momentum loss that the observed systems experienced. In order to do this, we estimated the initial masses and the period of the progenitors to the WR+O binaries. In this section, we describe the methods we used to estimate these initial properties of WR+O binaries. We then discuss how we used these initial properties to determine the mass-transfer efficiency and angular momentum loss of the binary systems, and we determine whether the systems experienced \revtwo{C}ase A or \revtwo{C}ase B mass transfer. 

\subsection{Initial masses}\label{initial_masses}

The observed masses of the WR and O star can be used to estimate the initial mass of the WR star in the binary. The way in which this is done differs between \revtwo{C}ase A and \revtwo{C}ase B. For \revtwo{C}ase B, this can be done based on the relation between the core mass and the initial mass, as found by \cite{Wellstein_1999}. As a WR star consists of just the core of the progenitor star, we can use this relation. \cite{Petrovic2005} made a linear fit of this relation, which resulted in equation \eqref{M_i_B},

\begin{equation}
    \label{M_i_B}
    M_{\textit{1},i,B} = \frac{M_{\textit{WR}} + 4.92}{0.53},
\end{equation}
\rev{where $M_{\textit{WR}}$ is the mass of the WR star, and $M_{\textit{1},i,B}$ is its progenitor mass for \revtwo{C}ase B.}

For \revtwo{C}ase A, we used figure 7 from \cite{Shao2016} to derive a similar relation. In this figure, they displayed the relation between the current WR mass and the initial mass of that star for different initial mass ratios \rev{and initial periods. We fit a straight line that traced the lower boundary of the majority of possible masses (i.e. those on the shortest orbital periods)}. This resulted in the relation between the current WR mass and the initial primary mass for \revtwo{C}ase A mass transfer that is found in equation \eqref{M_i_A}. Since this relation is approximately a lower limit, the outcome for $M_{\textit{1},i,A}$ \rev{based on a measured WR mass is an upper limit}. This allowed us to find the (approximate) initial primary mass for \revtwo{C}ase A,

  \begin{equation}
    \label{M_i_A}
    M_{\textit{1},i,A} = \frac{M_{\textit{WR}} + 4.86}{0.41},
  \end{equation}
\rev{where  $M_{\textit{1},i,A}$ is the maximum progenitor mass for \revtwo{C}ase A.}

With the initial mass of the WR star, we set some constraints on the parameter space of the secondary mass. First, the current WR star has to have been the donor star for it to have lost its envelope. Since the more massive star is the donor star, the initial mass of the current O star cannot be higher than the initial WR mass. Another way to derive an upper limit is by using the total mass of the system. Since we assumed that the O star has accreted mass, or at least not lost mass, we state that the initial O-star mass $M_{O,i} \leq M_{O,f}$, the final O-star mass. From these two limits, the lower of the two becomes the final upper limit for $M_{2,i}$. We also set a lower limit to obtain a better restricted parameter space. To do this, we used the knowledge that the total mass of the system cannot have grown, but only decreased or stayed the same. Thus, the sum of the initial masses $M_{1,i} + M_{2,i} \geq M_1 + M_2 = M_T$, the current total mass, which can be rewritten to $M_{2,i} \geq M_T - M_{1,i}$. This gives the lower limit for the initial mass of the companion. If the difference is especially large, this might become negative, in which case, we used zero as a lower bound instead. Together with the upper limit, we thus determined the range of the initial secondary mass for each of our systems.

The change in both the donor mass and the accretor mass can be expressed in terms of the accretion efficiency $\beta$ as $\Delta M_a = - \beta \Delta M_d$. This can also be expressed in terms of the total mass $M_T$ and the mass ratio $q=M_a/M_d$ \citep{Soberman_1997} as follows:
\begin{equation}
    \frac{M_T}{M_{T,i}} = \left(\frac{1+q}{1+q_i}\right)\left(\frac{1+\beta q_i}{1+\beta q}\right).
    \label{mt/mt0}
\end{equation}

\subsection{Initial period and mass-transfer type}

We cannot really estimate the initial period. We therefore considered many possible periods. Using the estimated initial masses for \revtwo{C}ase A and \revtwo{C}ase B, we then determined which periods lead to \revtwo{C}ase A and which to \revtwo{C}ase B mass transfer. To do this, we used the fact that mass transfer starts when a star fills its Roche \rev{lobe} $R_L$. More specifically, we used the fact that a star needs to have exhausted the hydrogen in its core before the onset of mass transfer for the mass transfer to be considered \revtwo{C}ase B. When we knew the initial mass of the donor, we determined its radius at the end of the main sequence, also called the terminal age main sequence (TAMS). This gives the condition $R_L \geq R_{d,\text{TAMS}}$ for \revtwo{C}ase B mass transfer. When the Roche lobe is smaller than the donor radius at TAMS, this indicates that Roche-lobe overflow (RLOF) has already occurred before, and the mass transfer is \revtwo{C}ase A. 

We related $R_L$ to the period $P$ and the mass ratio of the binary and the mass of the donor star $M_d$. To do this, we made use of the relation found by \cite{Eggleton1983} for \rev{$0.1 < Q < 10$, where $Q$ again is the mass ratio, but inverted: $M_d/M_a$},
\begin{equation}
    \label{r_l-approx}
    \frac{R_L}{a} \approx \frac{0.44 Q^{1/3}}{(1+Q)^{1/5}} .
\end{equation}
With Kepler's third law, 
\begin{equation}
    \label{p-tams}
    P = 9.859 \cdot (1+Q)^{-1/5}M_d^{-1/2}R_L^{3/2}\left(\frac{G}{4\pi^2}\right)^{-1/2} \cdot 10^{-7} \text{d}.
\end{equation}
The dependence on the mass ratio $Q$ has become small enough here, and we set $Q \approx 1$ to simplify the expression, resulting in
\begin{equation}
    P \propto M_d^{-1/2} R_L^{3/2}.
    \label{P-R-rel}
\end{equation}
\rev{With this equation, we determined the period boundary between \revtwo{C}ase A and \revtwo{C}ase B for each initial mass. We set it to the period where  $R_L = R_{d,\text{TAMS}}$.}

To derive the donor mass and radius at the relevant points of the donor evolution, we ran a simulation of the star with the code called modules for experiments in stellar astrophysics (MESA) \citep[and previous publications]{Paxton2018}. We used the inlists provided by \cite{Klencki2020}. These inlists were designed to simulate single stars from the beginning of the MS to the point in their evolution at which carbon in the core was exhausted. Even though we studied binaries, it is a safe assumption that the stars behave roughly like single stars until one of them fills their Roche lobe and mass transfer starts. Since we only studied the star up until the point at which it is possible for \revtwo{C}ase B mass transfer to take place, we only ran the simulations until the hydrogen in the core was exhausted \revtwo{(defined here as a hydrogen mass fraction in the core lower than $10^{-4}$)}.

We only made slight alterations to the inlists provided by \cite{Klencki2020}. We changed the initial mass and set the metallicity to $Z = Z_{\odot}$. We first created pre-\reveditor{zero age main-sequence} (ZAMS) models for each \revtwo{C}ase A and \revtwo{C}ase B initial primary mass using one set of inlists. Their outcome was used to simulate the stars from the ZAMS until the helium in their core was exhausted. This gave us the parameters we searched for to determine the boundaries for the period for \revtwo{C}ase A and \revtwo{C}ase B mass transfer. 

However, because solar metallicity might not be the correct metallicity for the systems, we repeated this process, but with $Z = 0.5 Z_{\odot}$. We did not study these outcomes in detail, but we briefly discuss some effects of this change in metallicity in section \ref{discussion}.

For more massive stars, the radius expands strongly at the end of the main sequence and obfuscates the boundary between \revtwo{C}ase A and \revtwo{C}ase B. In order to take this into account, we included the end of the main sequence in the calculations assuming \revtwo{C}ase A and \revtwo{C}ase B mass transfer.

\subsection{Accretion efficiency and angular momentum loss}\label{desired_information}

We sampled the parameter space of the initial periods and using 200 data points within the limits for the secondary initial masses and the initial periods, with an upper boundary of $10^3$ days for the period. \rev{In the minimum period, a ZAMS star would fill its Roche lobe.} The values for both parameters were spaced out evenly, linear for the masses and logarithmic for the periods. For each of the initial period and mass estimates, we derived the the accretion efficiency $\beta$ and the loss of angular momentum in terms of the \revtwo{} specific angular momentum loss per unit mass \revtwo{of the binary},
\begin{equation}
  \gamma  = \left(\frac{\Delta J_{\rm lost}}{\Delta M_{\rm lost}}\right)  \left(\frac{M}{J}\right) .  
\end{equation}
An equation for $\beta$ is fairly straightforward using equation \ref{mt/mt0}. After multiplying both sides by $(1 + \beta q)$ and rearranging the terms, we obtained an equation where it is only dependent on the initial and current masses of the system,
\begin{equation}
    \label{beta-eq}
    \beta = \left(\frac{1+q}{1+q_i} - \frac{M_T}{M_{T,i}}\right) \left(q\frac{M_T}{M_{T,i}} - q_i \frac{1+q}{1+q_i}\right)^{-1}.
\end{equation}

We also related the angular momentum loss $\gamma$ to the periods and masses of the WR+O binary and its possible progenitors. To do this, we used two equations, Kepler's third law to relate $P/P_i$, and  $M_T/M_{T,i}$ to the change in semi-major axis $a/a_i$. This change is related to $\gamma$ under the assumption that $\gamma$ is constant over time \citep{1994A&A...288..475P},
\begin{equation}
    \label{a/a0}
    \frac{a}{a_i} = \left(\frac{M_d}{M_{d,i}}\frac{M_a}{M_{a,i}}\right)^{-2} \left(\frac{M_T}{M_{T,i}}\right)^{2\gamma+1}.
\end{equation}

This was then equated to the relation for $a/a_i$ from Kepler's third law and was rewritten to give an equation for $\gamma$ that relates the period and masses of a WR+O binary and its prior,
\begin{eqnarray}
   \left(\frac{M_d}{M_{d,i}}\frac{M_a}{M_{a,i}}\right)^{-2} \left(\frac{M_T}{M_{T,i}}\right)^{2\gamma+1} &=& \left(\frac{M_T}{M_{T,i}}\right)^{1/3} \left(\frac{P}{P_i}\right)^{2/3} \\
   \left(\frac{M_T}{M_{T,i}}\right)^{2\gamma+\frac{2}{3}} &=& \left(\frac{P}{P_i}\right)^{2/3} \left(\frac{M_d}{M_{d,i}}\frac{M_a}{M_{a,i}}\right)^{2} \\
\end{eqnarray},
so that
\begin{equation}
    \label{gamma-eq}
    \gamma = \log_{M_T/M_{T,i}}\left(\left(\frac{P}{P_i}\right)^{1/3}\frac{M_1 M_2}{M_{1,i}M_{2,i}}\right) - \frac{1}{3}.
\end{equation}

\subsection{Limiting possible solutions}\label{set_lims}
The size of the grid is quite large, so that it can be useful to limit the part of the grid in which we are interested. We therefore studied this in more detail. We first disregarded all points where $\gamma < 0$ since this would mean that the system had gained angular momentum in some way. We also used an upper limit of $\gamma < 5$ since values higher than this limit are probably nonphysical. This is clear from considering mass loss through a circumbinary ring, as in \cite{Soberman_1997},
\begin{equation}
    \gamma = \frac{M_T^2}{M_d M_a}\cdot \frac{a_r}{a}^{1/2}
    \label{ring}.
\end{equation}
Here, $a_r$ is the radius of the ring.
For $\gamma = 3$ and $M_1=20M_{\odot}$ and $M_2=10M_{\odot}$, we would obtaine a radius of the ring $a_r$ that is four times as large as the radius of the binary. At some point, the question is whether this is still a reasonable radius for a circumbinary disk and if  is it still circumbinary, or just a disk. For $\gamma = 5$, this radius becomes even larger: $a_r \approx 11.1\cdot a$. A circumbinary disk with such a radius is not physically plausible, and we therefore used this value for $\gamma$ as an upper limit. We only consider points where $0 < \gamma < 5$ and deemed high values ($\gamma > 3$) unlikely.

The critical $q$ values further limit the parameter space. For $q > q_{\mathrm{crit}}$, mass transfer becomes unstable. When this occurs, we do not expect the binary to have become a WR+O binary. We therefore also disregarded the areas where $M_{1,i}/M_{2,i} > q_{\mathrm{crit}}$. However, the exact value is hard to determine because it depends on many variables, such as stellar radius and mass, but also on properties, such as efficiency and the case of the mass transfer. \cite{Ge2015} determined $q_{\mathrm{crit}}$ values for a range of different scenarios for conservative mass transfer. They reported that the critical $q$ value for a $16 M_{\odot}$ primary star ranges from $1.793$ to $8.646$. We do not expect the systems we studied to have undergone conservative mass transfer, and we therefore did not use this range for $q_{\mathrm{crit}}$. This shows, however, that there might not be a single correct value for this critical $q$, but instead a range for \revtwo{C}ase A and \revtwo{C}ase B. We disregarded results that lay outside the strictest boundary, but we took the part into account that was in between this and the most forgiving boundary. 

From the literature, the following ranges are satisfactory for either \revtwo{C}ase A or \revtwo{C}ase B mass transfer \citep{GallegosGarcia2022,Klencki2021},
\begin{eqnarray*}
    1.6 &\leq& q_{\mathrm{crit, A}} \leq 3\\ 
    4 &\leq& q_{\mathrm{crit, B}} \leq 10
\end{eqnarray*}

Finally, detailed studies \citep[e.g.][]{2022A&A...659A..98S} reported that very close binaries lead to a merger of the two stars, regardless of their mass ratio. The limiting period is highly uncertain and depends on assumptions, but the lower range of \revtwo{C}ase A periods is very likely ruled out as well.

\subsection{Most likely progenitors}
\begin{figure}
    \centering
    \includegraphics[width=0.45\textwidth]{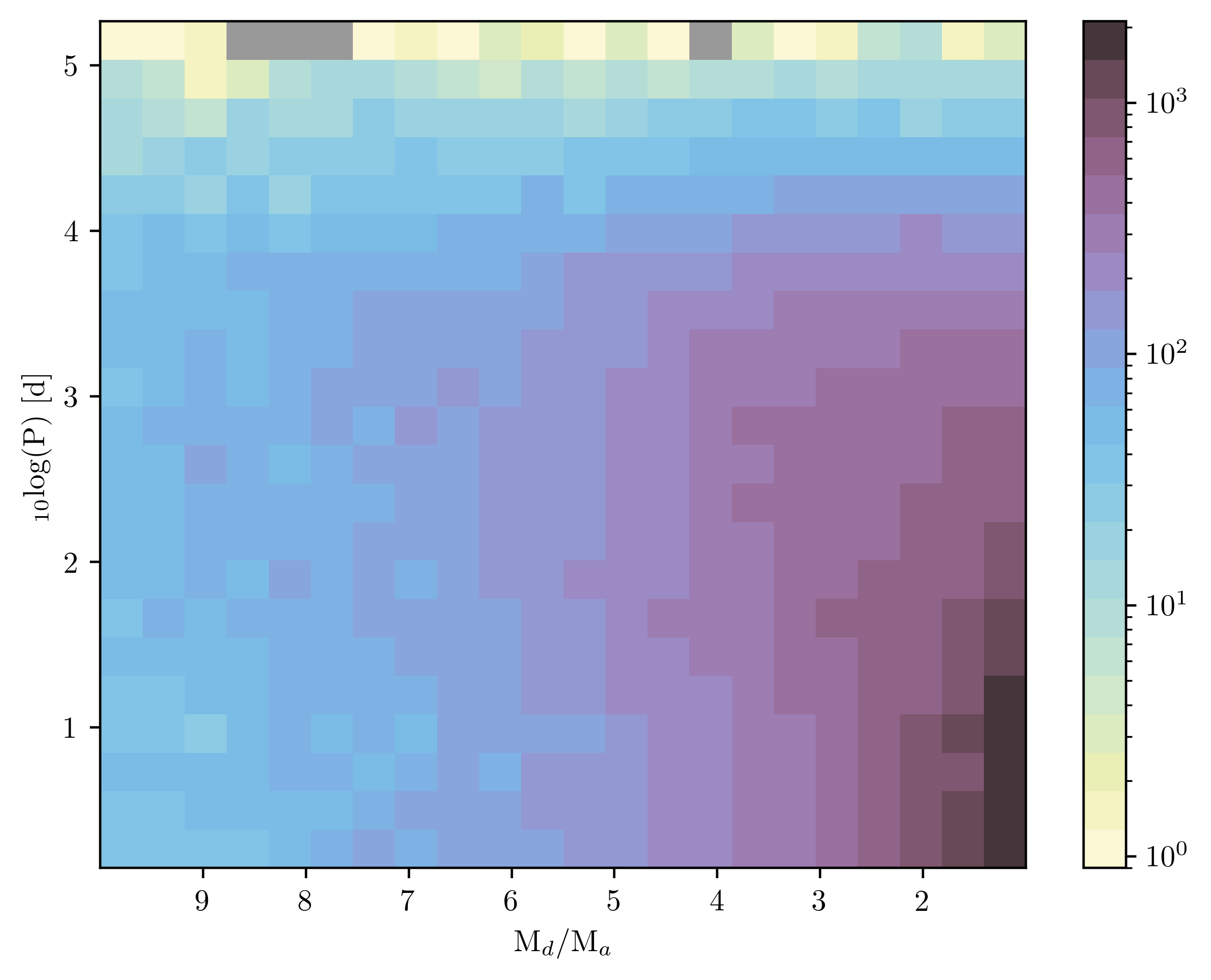}
    \caption{Period and mass ratio distribution for a sample of $10^5$ O+MS systems where $16M_{\odot}<M_1<150M_{\odot}$ and $q>1$, based on the distributions found by \cite{Moe2017}. \rev{The colour bar gives the number of systems per cell.}}
    \label{16_150}
\end{figure}

We also took prior information into account about the most common properties of binary systems similar to those of the WR+O star progenitors. The properties we considered are the mass ratio and the period. To obtain this information, we used the COSMIC code by \cite{cosmic} and the results of \cite{Moe2017} to create a sample of O+MS binaries with a size of $10^5$ and $M_{1,min} = 16$. This was to ensure that the primary would be an O star and that the sample was large enough. The random seed of the sample was 20. Since these types of binaries are expected to be the progenitors of the WR+O binaries we studied, we used this distribution to determine the likelihood of the different scenarios we might find for the progenitor evolution, such as the initial secondary mass. To do this, we studied the distribution of the O+MS mass ratios and periods. In figure \ref{16_150} we show a 2D histogram of the period and mass ratio of the sample. In the sample, we used a range for the primary mass ($16 M_{\odot}<M_1<150 M_{\odot}$) such that the initial primary star was an O star and $q>1$ because the O star was to be the donor and therefore had the higher initial stellar mass.

The distribution shows that the closer to $q = 1$, the more binaries, \rev{in particular for the systems with periods shorter than several dozen days}. For periods, there is a clear preference for lower periods of $\log P \lesssim 4$, and slightly so for $\log P \lesssim 2$. Following these results, we therefore expect the progenitors of the WR+O binaries to most likely have a mass ratio close to one and an orbital period that is not too high. These are the only possible values for the progenitors, but shows which values are more likely than others to have occurred. We used this result to determine the likelihood of certain scenarios we might find in our results. Scenarios with lower $q$ are more likely based on this O+MS distribution.

Finally, we determined for this distribution which systems we expect to undergo \revtwo{C}ase A mass transfer and for which systems we do not expect this \rev{based on our calculated period boundary between \revtwo{C}ase A and B}. About one-third of the O+MS systems (36,44\%) is expected to undergo \revtwo{C}ase A mass transfer, and the initial periods of the other two-thirds are too long for them to undergo \revtwo{C}ase A mass transfer. If all O+MS binaries have an equal likelihood to become a WR+O binary and the \rev{ observed WR+O binaries we studied are not too severely biased,} we would expect most of them not to have undergone \revtwo{C}ase A mass transfer.

\section{Objects}\label{Objects}

The systems we studied were all taken from the VIIth catalogue of Galactic Wolf-Rayet stars by \cite{vanderHucht2001}. This catalogue contains all the WR stars that are in the Milky Way. This iteration contains a total of 226 WR stars. We filtered the WR stars in the catalogue using the following criteria:
\begin{itemize}
    \item The WR star is in a binary.
    \item Its companion is an O star.
    \item The orbital period of the system is known.
    \item The primary and secondary mass are known.
\end{itemize}

In addition to the WR star being in a binary, its companion was to be an O star since these systems are likely progenitors of \reveditor{double black hole} binaries. To be able to usefully determine the remaining WR+O systems that are in the \reveditor{catalogue}, some additional information was required: the masses of the two stars and their orbital period. However, the \reveditor{catalogue} does not contain information on the stellar masses \rev{for all binaries}. We therefore had to obtain these masses from other sources, \rev{and these are indicated in the table}.

This resulted in a sample of \rev{24} WR+O binaries in the Galaxy. Initially, we worked with this sample. However, during the simulations of the WR progenitor stars, the masses of some of the progenitors were found to be too high to do this properly. These were the progenitors of the three systems with $M_{\textit{WR}} \gtrsim 45 M_{\odot}$, \object{WR22}, \object{WR47}, and \object{WR141}. The other \rev{21} had $M_{\textit{WR}} \lesssim 25 M_{\odot}$. The high WR masses of the three systems resulted in initial primary masses of $M_{1,i} \gtrsim 100 M_{\odot}$. These initial masses proved to be too high to subsequently run proper simulations of the donor star from ZAMS to TAMS. Therefore, we discarded these three systems from our selection. \rev{WR35a was found to be a binary after the latest version of the \cite{vanderHucht2001} by \citet{2014A&A...562A..13G}.}

An overview of the final sample consisting of \rev{21} WR+O binaries and some of their properties is listed in table \ref{objects-tab}. When no source is specified, the data were taken from \cite{vanderHucht2001}. The mass ratio was determined using the masses, and the Roche-lobe radius of the WR star $R_L$ was determined using Eq.~\ref{p-tams}.

\begin{center}
\begin{table*}[bt]
    \caption{Table of WR+O binaries in sample}
    \centering
    \begin{adjustbox}{scale=1, center}
    \begin{tabular}{cccccccc}
        \hline \hline														
        WR\#	&	Spectral type	&	Binary status	&	$P$ (d)	&	$M_{\textit{WR}}$ ($M_{\odot}$)	&	$M_O$ ($M_{\odot}$)&	$q$	&	$R_L$ ($R_{\odot}$)	\\ \hline
        9 	&	\object{WC5+O7}  	&	SB2	&	14.3	&	9	&	32	&	0.28	&	23.45	\\ 
        11 	&	WC8+O7.5III-V  	&	SB2, VB	&	78.5	&	9.0 $\pm$ 0.6$^a$&	28.5 $\pm$ 1.1$^a$&	0.32	&	73.24	\\ 
        21 	&	WN5o+O4-6;  	&	SB2 	&	8.25	&	19	&	37	&	0.51	&	21.32	\\
        	&	WN5o+O7V 	&		&		&		&		&		&		\\ 
        30 	&	WC6+O6-8;  	&	SB2	&	18.8	&	16	&	34	&	0.47	&	34.73	\\
        	&	WC6+O7.5 	&		&		&		&		&		&		\\ 
        31 	&	WN4o+O8V  	&	SB2, VB	&	4.83	&	>11$^b$&	>24$^b$&	0.46	&	12.37	\\ 
        35a 	&	WN6 + O8.5 V  	&	SB2	&	41.9$^c$	&	19$^c$&	18$^c$&	1.1	&	45.1	\\ 
        42 	&	WC7+O7V  	&	SB2	&	7.89	&	14	&	23	&	0.61	&	18.84	\\ 
        48 	&	WC6(+O9.5/B0Iab);  	&	SB1, VB	&	19.14	&	18$^d$&	29$^d$&	0.62	&	37.03	\\
        	&	WC6+O6-7V 	&		&		&		&		&		&		\\ 
        62a 	&	WN4-5o; 	&	SB2	&	9.1	&	>21.5 $\pm$ 4.8	&	>38.7 $\pm$ 5.2	&	0.55	&	24.03	\\
        	&	WN5+O5.5-6;	&		&		&	(of > 22.6 $\pm$ 5)$e$	&	(of >42.0 $\pm$ 5.0)$^e$	&		&		\\
        	&	WN6o	&		&		&		&		&		&		\\ 
        68a 	&	WN6o+O5.5-6;	&	SB2	&	5.22	&	>15 $\pm$ 5$^f$	&	>30 $\pm$ 4$^f$	&	0.50	&	14.50	\\
        	&	WN6o  	&		&		&		&		&		&		\\ 
        79 	&	WC7+O5-8  	&	SB2, VB	&	8.89	&	11	&	29	&	0.38	&	18.45	\\ 
        97 	&	WN5b+O7  	&	SB2	&	12.6	&	2.3	&	4.1	&	0.56	&	14.04	\\ 
        113 	&	WC8d+O8-9IV  	&	SB2	&	29.7	&	13	&	27	&	0.48	&	44.00	\\ 
        127 	&	WN3b+O9.5V  	&	SB2	&	9.56	&	17	&	36	&	0.47	&	22.58	\\ 
        133 	&	WN5o+O9I  	&	SB2, VB	&	112.8	&	9.3 $\pm$ 1.6$^g$	&	22.6 $\pm$ 3.2$^g$	&	0.13	&	92.42	\\ 
        137 	&	WC7pd+O9;	&	SB2	&	4763.25	&	4.4 $\pm$ 1.5$^h$	&	20 $\pm$ 2$^h$	&	0.22	&	882.00	\\
        	&	WC7ed+O9  	&		&		&		&		&		&		\\ 
        139	&	WN5o+O6III-V  	&	SB2	&	4.2	&	9.3	&	28	&	0.33	&	10.53	\\ 
        140 	&	WC7pd+O4-5;  	&	SB2, VB;	&	2898.1	&	16 $\pm$ 3$^i$	&	41 $\pm$ 6$^i$	&	0.39	&	990.97	\\
        	&	WC7ed+O5	&	CWB	&		&		&		&		&		\\ 
        151 	&	WN4o+O5V  	&	SB2 	&	2.13	&	20	&	28	&	0.71	&	8.94	\\ 
        153 	&	WN6o/CE+O6I; 	&	SB2 + SB2	&	6.69 + 3.47	&	>6$^k$	&	>21$^k$	&	0.29	&	7.97	\\
        	&	WN6o/CE+O3-6+B0:I+B1:V-III  	&		&		&		&		&		&		\\ 
        155 	&	WN6o+O9II-Ib  	&	SB2	&	1.64	&	24	&	30	&	0.80	&	8.03	\\ \hline
    \end{tabular}
    \end{adjustbox}
    \tablefoot{All data were taken from \cite{vanderHucht2001}, and a: \citet{North2007}, b: \citet{Vanbeveren2020}, c: \citet{2014A&A...562A..13G}, d:  \citet{LenoirCraig2021}, e: \citet{Collado2013}, f: \citet{Collado2015}, g: \citet{Richardson2021}, h: \citet{Lefvre2005}, i: \citet{Fahed2011}, and k: \citet{Demers2002}}
    \label{objects-tab}
\end{table*}
\end{center}

\FloatBarrier
\section{Results}\label{Results}

We present the results of our study in figures for each object. As an example, we use WR9, which is shown in figure \ref{contour_WR9}. The main graph is divided into two parts, in which the upper half shows \revtwo{C}ase B and the lower half shows \revtwo{C}ase A. The initial mass of the WR star as estimated is given in a box for each of the two cases. The x-axis at the bottom shows initial secondary masses covering the allowed ranges (see Sect.~\ref{initial_masses}) for both of the cases. The y-axis shows the range of initial periods. The colour gradient within the parts for \revtwo{C}ase A and B displays the $\gamma$ values we found for each point in the grid of initial secondary masses and periods for \revtwo{C}ase A and \revtwo{C}ase B, and the colour bar on the side shows the legend. \rev{Every integer value of $\gamma$ is marked with a black line for clarity.} Values of $\gamma$ below zero (white) and above 5 (black) were ruled out. Some parts of the grid appear to be more shaded out than others. These shaded areas indicate the critical $q$ values. The parts without shading always experience stable mass transfer based on the $q_{\textit{crit}}$ range used. The partly shaded out parts could have experienced either stable or unstable mass transfer, depending on the exact value for $q_{\textit{crit}}$. The parts that are shaded out heavily always experience unstable mass transfer based on the critical $q$ values. For the part of the graph that displays the grid for \revtwo{C}ase B, \revtwo{we show the lower limit of $P_i$ for \revtwo{C}ase B mass transfer to occur as the dashed line. In the part below the dashed line, the shaded areas are different than for the rest of the graph. This represents late \revtwo{C}ase A mass transfer that  may behave more like \revtwo{C}ase B due to the composition of the donor. It is unclear whether the critical $q$ values would then also be those of Case B. Therefore, we (also) show the critical $q$ values of Case A. These do not correspond to the same possible values for the initial secondary mass $M_{2,i}$ of the lower, Case A, part of the graph because the initial mass of the primary $M_i$ is different, and therefore, the mass ratio $q$ is also different.}
Lastly, the smaller graph at the bottom displays the mass-transfer efficiency $\beta$ for both cases of mass transfer, depending only on the initial secondary mass and not on the initial period. This graph uses the same values for the x-axis as the other two graphs. The y-axis displays the value for $\beta$ for a certain mass ratio $q$.

For each observed system, we determined which of the two mass-transfer types is possible and the possible and likely values of $\beta$ and $\gamma$. We present the results according to the possibilities.

\subsection{Mass-transfer type}\label{cases-sec}

\subsubsection{Case A is more likely}

For many systems, we conclude that \revtwo{C}ase A is the most likely, and this is the largest part of our sample. Fourteen of the 20 systems we studied fall into this category based on our results: WR9, WR21, WR30, WR31, WR42, WR48, WR62a, WR68a, WR79, WR127, WR139, WR151, WR153, and WR155. 

As an example, we consider WR9 (figure \ref{contour_WR9}). The results for the other systems are provided in the appendix \ref{appendix}. For this system, \revtwo{C}ase B is not fully excluded, but to have experienced \revtwo{C}ase B, it would take on extremely high values ($\gamma \gtrsim 4$) which while not impossible are not plausible. A similar conclusion holds for the other systems in this category. Instead, the \revtwo{C}ase A results are possible in the sense that values of $\gamma$ in the expected range are found. For the late \revtwo{C}ase A results (top panel below the dashed line), solutions with plausible $\gamma$ values are possible for many of the systems, while for others, the $\gamma$ values are rather high again ($\gamma \gtrsim 3$)

Some limits are imposed on all systems by the critical $q$ values concerning which progenitors would have undergone stable \revtwo{C}ase A mass transfer. The degree to which these limits affect the possibilities differ for each system, but all systems have options towards the upper limit of the secondary mass, depending on which value for the critical $q$ is assumed. However, for about half of these systems \rev{stable \revtwo{C}ase A evolution is only possible if the higher critical $q$ value of 3 is true. These systems, WR31, WR42, WR48, WR62a, WR68a, WR151, and WR155, have the most different initial mass ratios, and all have rather short periods (but they overlap with the other systems). The progenitor mass we used for \revtwo{C}ase A is an upper limit. Lower progenitor masses increase the initial mass ratio and thus allow for more stable mass transfer. The expected progenitor distribution peaks towards shorter periods and more equal masses and places the a priori most likely progenitors also in the region where stable \revtwo{C}ase A mass transfer is possible.} In these scenarios, the mass-transfer efficiency most likely would have been very low. We discuss the mass-transfer efficiency for all systems in more detail in section \ref{eff-res}.  We repeat here that towards the bottom of the \revtwo{C}ase A box, the systems likely merge (see section~\ref{set_lims}).

In summary, based on the values for $\gamma$, we exclude \revtwo{C}ase B mass transfer as a plausible scenario for these 14 systems. Case A mass transfer with a low mass-transfer efficiency seems to be the most likely scenario for these binaries based on our results.

\begin{figure}
    \centering
    \includegraphics[width=0.45\textwidth]{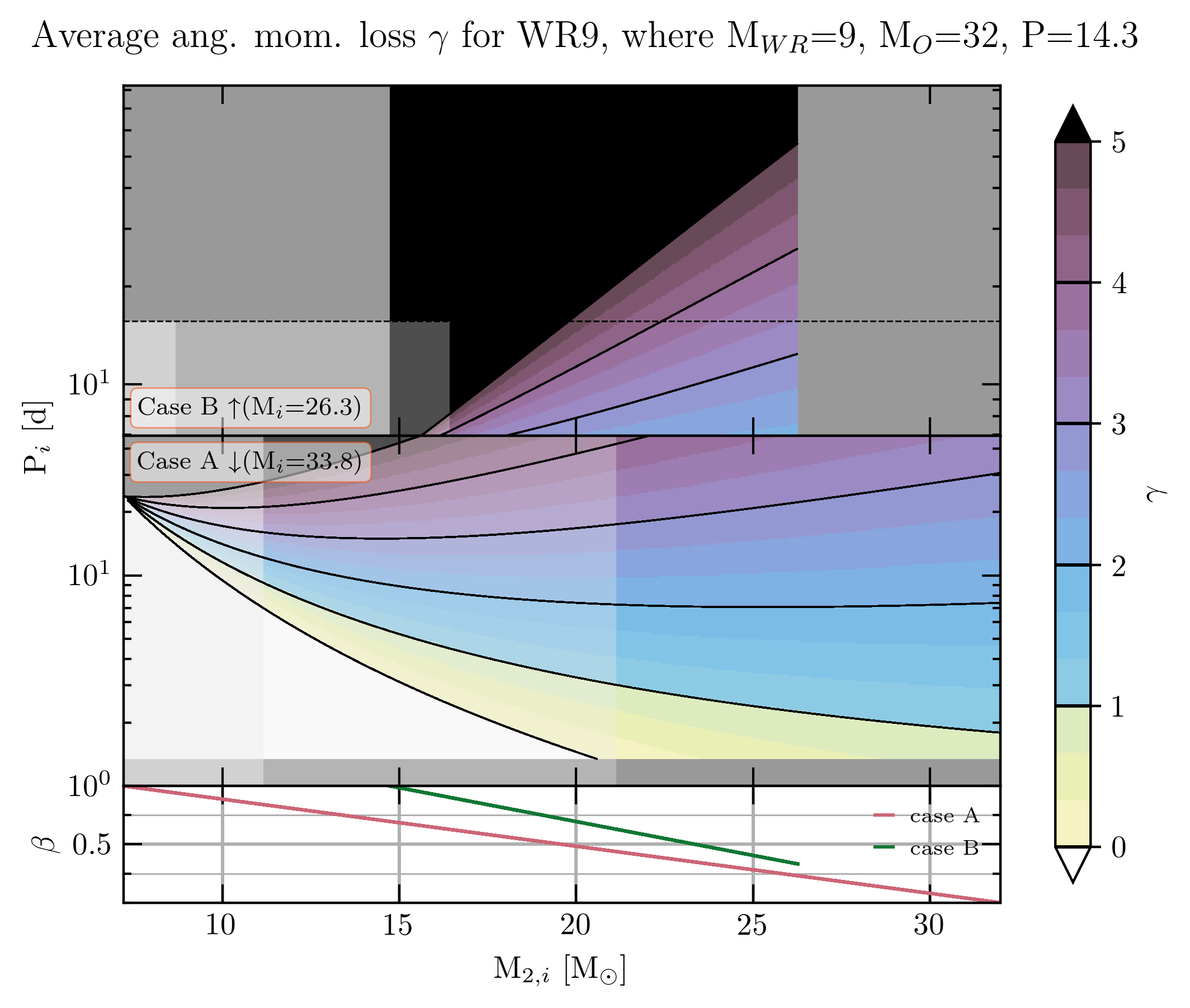}
    \caption{Resulting values for $\gamma$ (colour) and $\beta$ (lines in lowest graph) for the initial parameter space of WR9. \rev{Each integer value of $\gamma$ is marked with a solid line.} The shaded boxes indicate whether mass transfer will be stable. Parts without shading will be stable, lightly shaded parts can be either stable or unstable, depending on the exact value of $q_{\textit{crit}}$  The heaviest shaded parts will always experience unstable mass transfer. The dashed line indicates the period boundary for \revtwo{C}ase B. Dark grey shades indicate excluded mass ranges (see text).}
    \label{contour_WR9}
\end{figure}

\subsubsection{Case A and \revtwo{C}ase B are both possible}
For some systems in our sample, we were unable to determine the most likely cases of mass transfer because \revtwo{C}ase A and \revtwo{C}ase B both seem possible. Three systems in our sample are unclear: WR11, WR113, and WR133. \rev{WR35a is also marginally in this class because a small range of \revtwo{C}ase A is allowed. However, \revtwo{C}ase B seems more likely for this system (see below).} We examine the plot WR11 as an example (figure~\ref{contour_WR11}. The plots containing the other results are shown in the appendix \ref{appendix}.

For these three systems, for \revtwo{C}ase A and \revtwo{C}ase B there are many possibilities in which the mass transfer would have been stable, regardless of which value for $q_{\textit{crit}}$ is assumed. While this poses some limits for \revtwo{C}ase A, it still leaves enough possibilities. The values for $\gamma$ are also within reason ($\lesssim 3$) for \revtwo{C}ase A and large parts of the \revtwo{C}ase B grid. Finally, one case of mass transfer might be found to be more likely than the other based on a comparison of the results to the progenitor distribution. This distribution shows that low initial mass ratios are more likely with initial periods that are also short\rev{, although for the \revtwo{C}ase B region, the mass ratios are more equally distributed (see Fig.~\ref{16_150}).}. While this would shift the preference to a lower $\beta$ \rev{for \revtwo{C}ase A,} it does not impose any other limits for these three systems that would make either \revtwo{C}ase A or \revtwo{C}ase B very unlikely because even for low $q$ and $P$, there are still plausible scenarios for both cases of mass transfer to be found.

Our results have no further options to rule out any of the two cases of mass transfer. Therefore, these systems might have undergone either \revtwo{C}ase A or \revtwo{C}ase B mass transfer based on our results.

\begin{figure}
    \centering
    \includegraphics[width=0.45\textwidth]{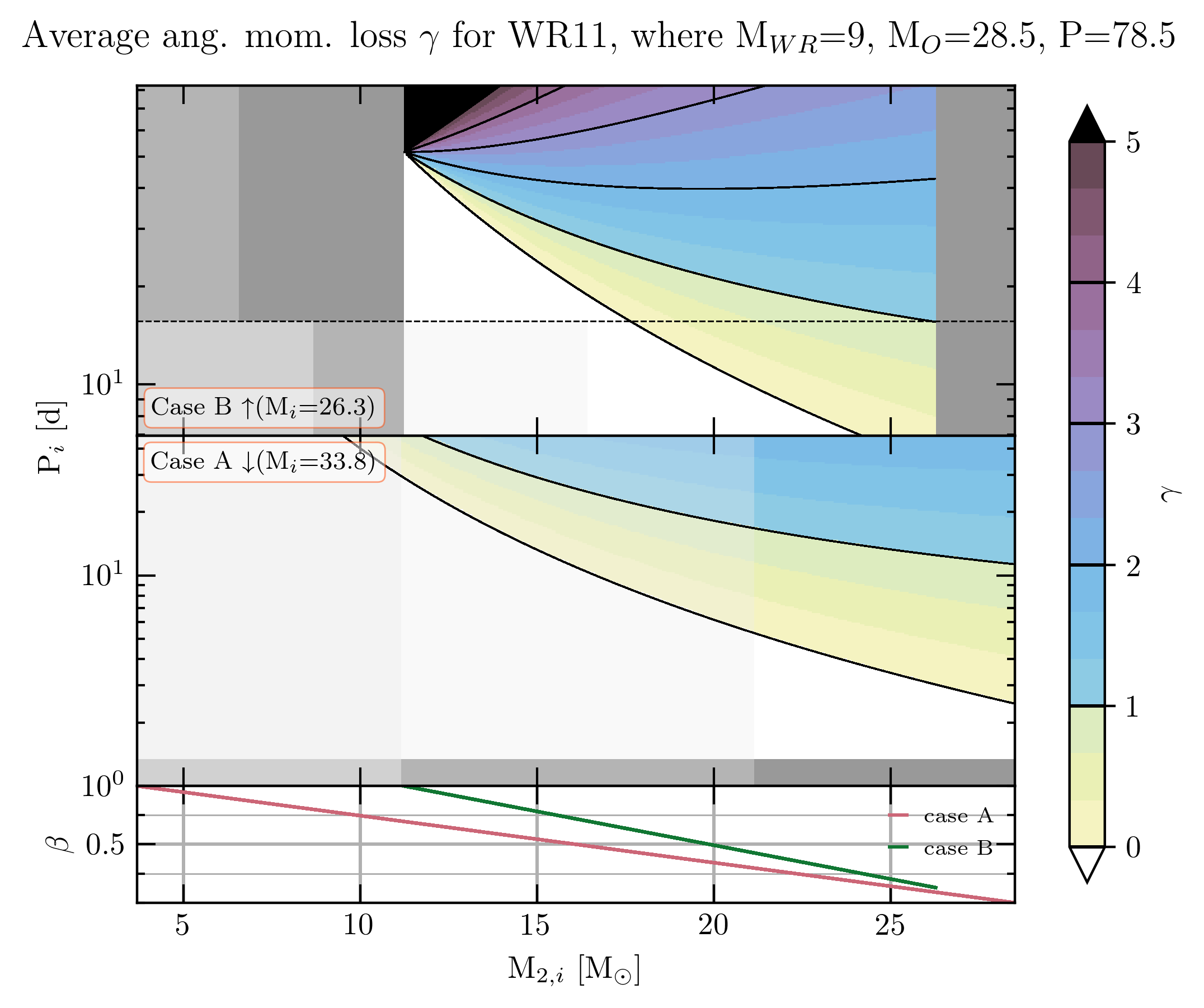}
    \caption{Same as figure \ref{contour_WR9}, but for WR11.}
    \label{contour_WR11}
\end{figure}

\subsubsection{Case A is not possible}
Instead of \revtwo{C}ase A being more likely, we also have 
three systems where based on our results we deem \revtwo{C}ase B mass transfer to be the most likely scenario. These systems are WR97\rev{, WR35a}, and WR140. 

\begin{figure}
    \centering
    \includegraphics[width=0.45\textwidth]{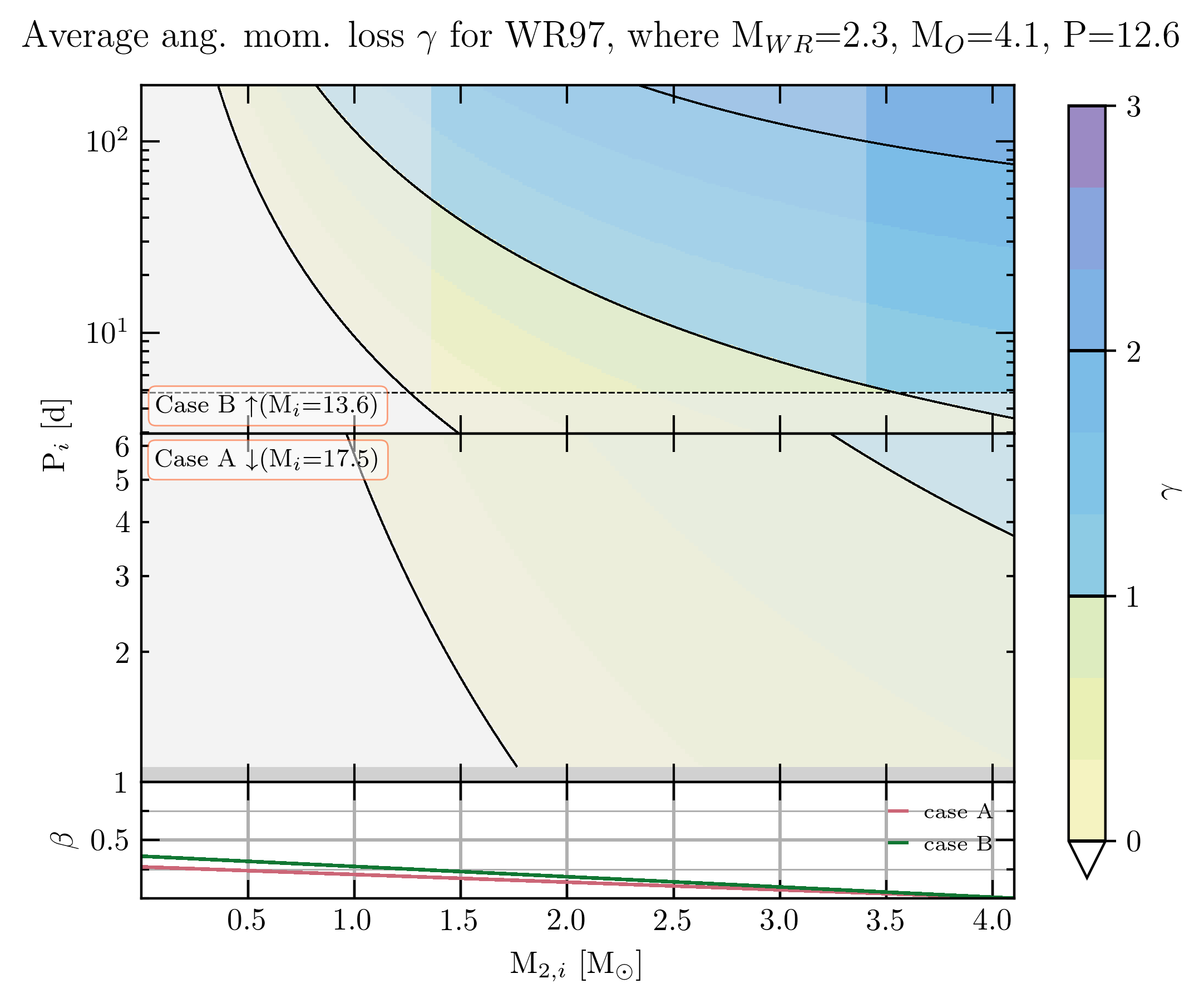}
    \caption{Same as figure \ref{contour_WR9}, but for WR97.}
    \label{contour_WR97}
\end{figure}

We followed the same procedure for $\gamma$ and $q_{\textit{crit}}$. For WR97 (figure~\ref{contour_WR97}), all possible \revtwo{C}ase A progenitors are predicted to undergo unstable mass transfer even by the most  least restraining $q_{\textit{crit}}$ value, and therefore, we excluded them as a possibility. For \revtwo{C}ase B, some limitations are also set by the critical $q$ values for this system, but in a part, mass transfer is still certainly stable with reasonable $\gamma$ values. WR97 is peculiar in the sense that the masses of the WR and the O star are quite low.

\rev{For WR35a (see the appendix \ref{appendix}), there is a small range of possibilities for \revtwo{C}ase A, but \revtwo{C}ase B is more likely. Due to the high current mass ratio, only low values of $\beta$ are possible.}

\begin{figure}
    \centering
    \includegraphics[width=0.45\textwidth]{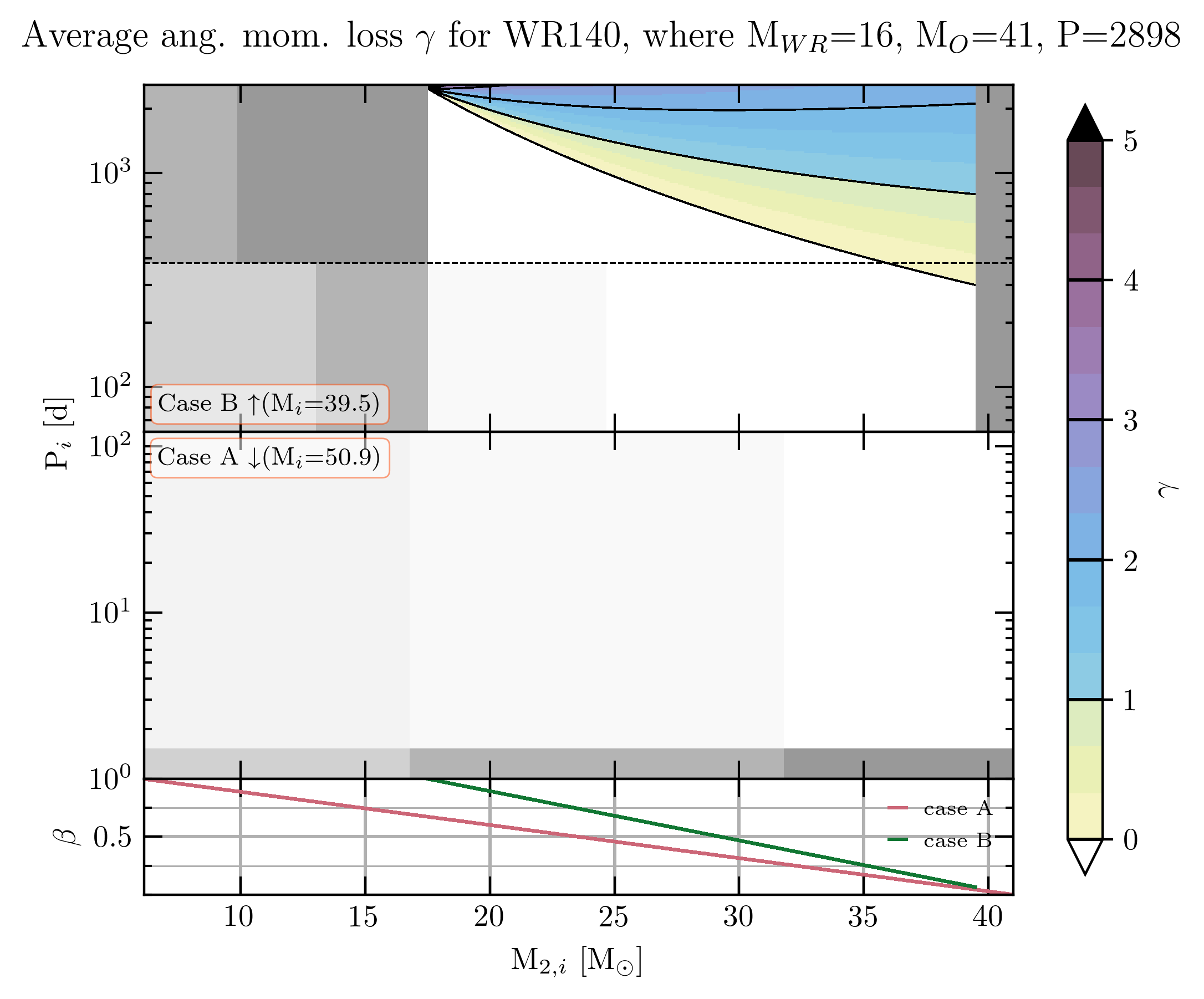}
    \vspace*{-4mm}
    \caption{Same as \ref{contour_WR9}, but for WR140.}
    \label{contour_WR140}
\end{figure}

For WR140 (figure~\ref{contour_WR140}), \rev{there are also options for \revtwo{C}ase B, but} the situation is different. For \revtwo{C}ase A, there are no solutions with positive $\gamma$. This is a consequence of the very wide period of the system, which means that it could even have been Case C or might have avoided mass transfer altogether.

\begin{figure}
    \centering
    \includegraphics[width=0.45\textwidth]{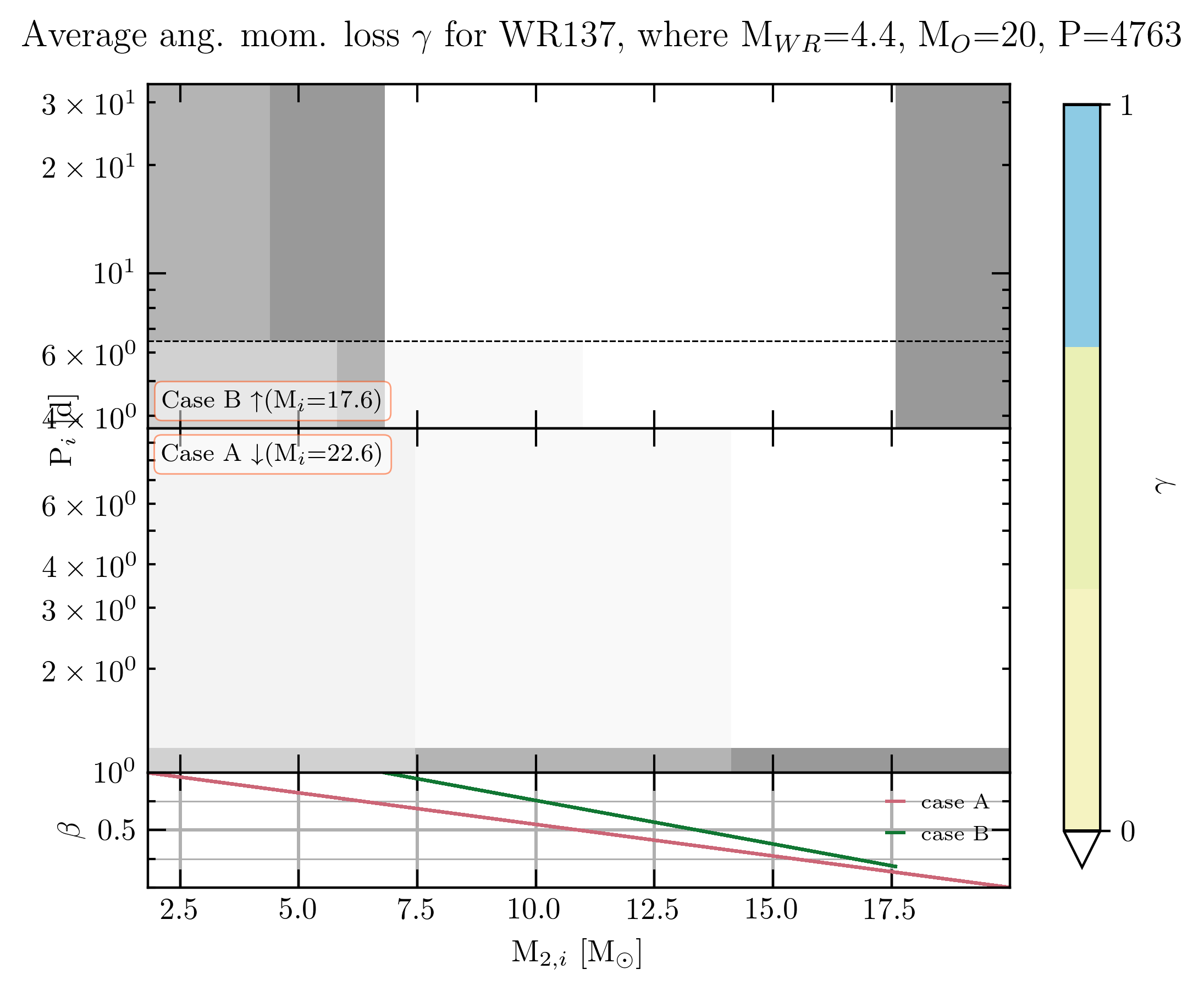}
    \caption{Same as figure \ref{contour_WR9}, but for WR137.}
    \label{contour_WR137}
\end{figure}

Finally, for WR 137 (figure~\ref{contour_WR137}), neither \revtwo{C}ase A nor \revtwo{C}ase B are likely scenarios. There are no solutions because the period of the system is very wide. As in the case of WR 140, mass transfer may have been avoided.

\subsection{Mass-transfer efficiency and angular momentum loss}\label{eff-res}
The possible mass-transfer efficiencies of a system can be determined by using the $q_{\textit{crit}}$ values for each case of mass transfer as an upper limit. This was done by determining the expected $\beta$ for each system for the two cases of mass transfer as a function of the initial secondary mass. This was then combined with the critical $q$ values for \revtwo{C}ase A and \revtwo{C}ase B to determine an upper boundary for each system. However, because there is a range of values for $q_{\textit{crit}}$ instead of one value, a range for $\beta_{\mathrm{max}}$ followed. For the lower and upper boundary of $q_{\textit{crit}}$, we found the matching mass-transfer efficiency and used this as an upper limit. For \revtwo{C}ase A, this corresponds to 1.6 < $q_{\mathrm{crit,A}}$ < 3 and for \revtwo{C}ase B to 4 < $q_{\mathrm{crit,B}}$ < 10.  An overview of the matching upper limits for $\beta$ for each binary is given in table \ref{beta-qlims}. We only list the limits for plausible cases of mass transfer. More detail of this determination was described in previous sections.

\begin{table}
    \caption{Table of upper limits for $\beta$ as determined by $q_{\textit{crit}}$}
    \centering
    \begin{tabular}{|c|c|c|c|c|}
    \hline
    WR\# & $q_{\mathrm{crit,A}} = 3$ & $q_{\mathrm{crit,A}} = 1.6$ & $q_{\mathrm{crit,B}} = 10$ & $q_{\mathrm{crit,B}} = 4$ \\ \hline
    9   & 0.84  & 0.44  &   &   \\ \hline
    11  & 0.70  & 0.30  & 1.00  & 1.00  \\ \hline
    21  & 0.46  & 0.01  &   &   \\ \hline
    30  & 0.49  & 0.06  &   &   \\ \hline
    31  & 0.41  & 0.00  &   &   \\ \hline
    35a & 0.02  & 0.00  &  0.58  & 0.33   \\ \hline
    42  & 0.25  & 0.00  &   &   \\ \hline
    48  & 0.28  & 0.00  &   &   \\ \hline
    62a & 0.43  & 0.00  &   &   \\ \hline
    68a & 0.42  & 0.00  &   &   \\ \hline
    79  & 0.59  & 0.18  &   &   \\ \hline
    97  &       &       & 0.24  & 0.06  \\ \hline
    113 & 0.41  & 0.00  & 1.00  & 0.89  \\ \hline
    127 & 0.51  & 0.07  &   &   \\ \hline
    133 & 0.45  & 0.04  & 1.00  & 0.90  \\ \hline
    137 &   &   &   &   \\ \hline
    139 & 0.66  & 0.26  &   &   \\ \hline
    140 &   &   & 1.00  & 1.00  \\ \hline
    151 & 0.20  & 0.00  &   &   \\ \hline
    153 & 0.60  & 0.22  &   &   \\ \hline
    155 & 0.15  & 0.00  &   &   \\ \hline
    \end{tabular}
    \tablefoot{The limits are only displayed for plausible case(s) of mass transfer. Depending on which value for $q_{\textit{crit}}$ is chosen, the upper limit of $\beta$ is different.  $\beta_{\mathrm{max}}$ certainly falls somewhere between the two values for each case that is displayed in this table. Other parameters such as $\gamma$ were not taken into account for these limits.}
    \label{beta-qlims}
\end{table}

For \revtwo{C}ase A, we expected to find a higher mass-transfer efficiency with values ranging from $\sim0.1-0.7$, while for \revtwo{C}ase B, we expected to find lower values of $\lesssim 0.25$ for the efficiency \citep[e.g.][]{2022A&A...659A..98S,Shao2016,2007A&A...467.1181D}. For the \revtwo{C}ase A results, the efficiency is a combination of a low efficiency of the fast mass-transfer phase and a higher efficiency in the slow phase \citep[see][]{2022A&A...659A..98S}. However, the values we found for $\beta_{\mathrm{max}}$ for the systems show something different. Firstly, the upper limits we found for \revtwo{C}ase B are \rev{often} very high ($\gtrsim 0.9$), regardless of which value for $q_{\textit{crit}}$ was taken. This seems to contradict the expectations we had of finding $\beta \lesssim 0.25$ for \revtwo{C}ase B for one of the systems where we expected \revtwo{C}ase B to be the most likely scenario, WR140.  However, a high maximum value does not automatically have to mean that this is also the efficiency that the mass transfer had.  For WR97 \rev{and WR35a}, we found low maximum values of $\beta$.
%If we use the expectations for the O+MS distribution the likelihood of lower efficiencies increases as those are associated with lower initial values for $q$. Therefore it is still very likely for these \revtwo{C}ase B systems to have undergone highly non-conservative mass transfer, although based on our results we cannot exclude scenarios where $\beta \gtrsim 0.25$.

For \revtwo{C}ase A, we find very different values for $\beta_{\mathrm{max}}$. The majority of the systems we studied has $\beta_{\mathrm{max}} < 0.5$ even for the least restraining value of $q_{\textit{crit}}$. The strictest boundary leaves no systems with $\beta_{\mathrm{max}} > 0.5$ and roughly half of the systems with $\beta_{\mathrm{max}} \sim 0$, which would mean that \revtwo{C}ase A mass transfer is not the most likely scenario for these systems, combined with these critical $q$ values. These results also disagree with the expectations we had, just as for \revtwo{C}ase B. We expected values of $\beta_{\mathrm{max}} \sim 0.1-0.7$. While these possibilities again are not excluded by our results, a lower maximum seems the more likely scenario based on our results. This is supported even more strongly when we again take the progenitor distribution and the higher likelihood of lower $q$ into account, and with this, lower values for $\beta$.

Therefore, we find that \rev{at least} for \revtwo{C}ase A mass transfer, the mass-transfer efficiency was most likely low due to the progenitor distributions. This is even more so because of the limitations posed by $q_{\textit{crit}}$.

The most likely values for the parameter $\gamma$ are often fixed to a certain value, or it is assumed that the mass that is lost takes away the angular momentum of the accretor. For binaries with initial mass ratios not far from unity, the latter implies values of $\gamma$ around one. The transition from $\gamma$ from below to above unity is visible in most of the figures. It implies that values around unity are indeed possible. However, the low $\gamma$ values typically come from initially short periods that may very well lead to mergers, suggesting that higher $\gamma$ values \rev{above  one} are more likely. The figures in the appendix show that some systems (WR31, \rev{WR35a if it was \revtwo{C}ase B}, WR139, WR79, WR151, WR153, and WR155) only have solutions of $\gamma$ above one. \rev{Our initial masses of the WR progenitors are upper limits (see Sect.~\ref{desired_information}, and lower initial masses lead to somewhat lower values of $\gamma$ (Eq.~\ref{gamma-eq})}.

\section{Discussion}\label{discussion}
We discuss the assumptions we made and their possible influence on the results and conclusions. 
Firstly, we assumed that the companion star did not or could not have lost mass with respect to its initial mass. However, this is not entirely correct. Stars can lose mass through stellar winds. While the amount of mass that is lost in this way is not expected to be high, it does mean that it is possible for the secondary star to slightly decrease in mass with respect to its initial mass $M_{2,i}$.

Another caveat that has to do with the mass is the way in which the masses of the WR progenitors were determined for \revtwo{C}ase A and \revtwo{C}ase B mass transfer. We used a linear relation for each, as explained in \ref{Method}. These relations are an approximation. While the initial progenitor masses are close to the outcomes of these equations, they might also be slightly lower or higher than the masses we used for our results. \rev{However, by including part of the \revtwo{C}ase A parameter range in the top panels of the plots, we cover a significant fraction of this uncertainty.} Another factor for this uncertainty is that some of the WR \rev{and O-star} masses  are uncertain, as we showed in section \ref{Objects}. 
\rev{We did not systematically vary the masses, but the effect of the changes in the mass is as follows: Lower WR masses move the curves down in the figures, that is, they increase the values of $\gamma$. Increasing the O-star masses has the same effect. We therefore do not expect a specific bias in our results, except for the one mentioned above, which is due to our \revtwo{C}ase A progenitor masses being upper limits. }

We assumed solar metallicity for the systems that we studied, although we are not fully certain that this is accurate. We therefore also examined the effect of a lower metallicity on the most likely scenario for a system. The outcomes for $\beta$ and $\gamma$ for a certain combination of initial parameters are not altered because they do not depend on metallicity. However, the period boundaries are dependent on the radius of the star at certain points in its evolution, which does change with metallicity. The resulting core masses also depend on the metallicity.

We ran the simulations to determine these boundaries again, but for $Z = 0.5Z_{\odot}$. For these simulations,  the radius of the star at the TAMS is smaller for a lower $Z$. This radius is indicated with the dot for each track. Because of this, the values for $P_i$ also become lower.

\begin{figure}
    \centering
    \includegraphics[width=0.45\textwidth]{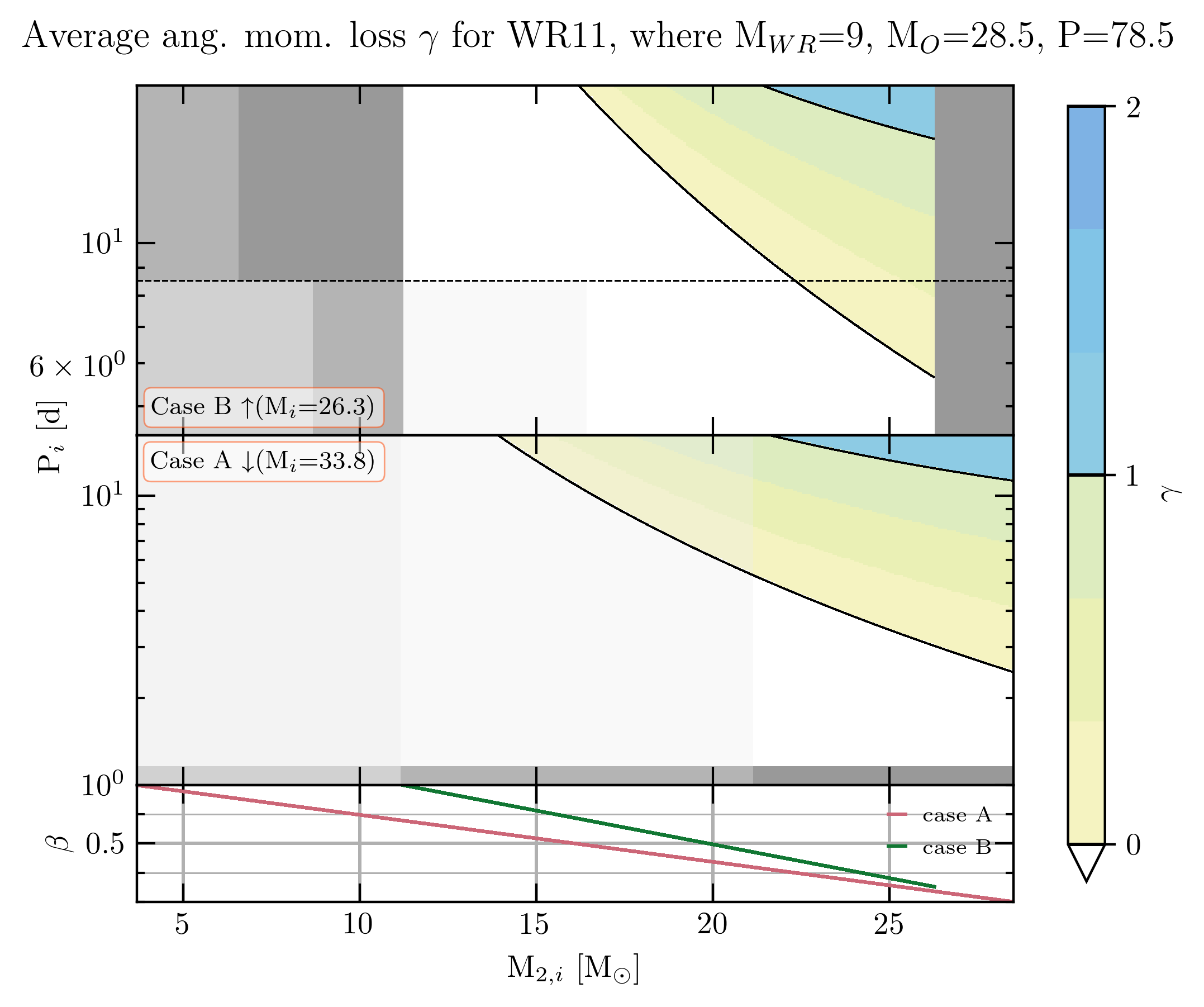}
    \caption{Same as figure \ref{contour_WR11}, but for a metallicity of $Z=0.5Z_{\odot}$.}
    \label{contour_WR11_lowz}
\end{figure}

This difference in evolution for our results is shown in a new figure for WR11 to illustrate the effect (figure~\ref{contour_WR11_lowz}. There are less possible scenarios for \revtwo{C}ases A and B. Maybe more interestingly, for \revtwo{C}ase B lower and more plausible values for $\gamma$ become possible within these new boundaries, whereas the possibilities for the initial period decrease drastically for \revtwo{C}ase B. \rev{At the same time, the conclusion that $\gamma > 1$  most likely changes as there now is an upper limit on $\gamma$ just above one.} Based on these results, a different metallicity might therefore influence the likelihood of the four scenarios we studied \rev{and our conclusion on the most likely value of $\gamma$}. It might be interesting to study this in more detail in future research.

Lastly, there is also the uncertainty on the exact value for the critical $q$ above which mass transfer becomes unstable. We used different values from the literature and all values in between as a possible range, but in reality, there is a single value of $q_{\textit{crit}}$ for each system that is able to determine whether the mass transfer will be unstable. Since it requires much information to accurately determine this, as reported for example by \cite{Ge2015}, it was not possible for us to do so as we did not have all of the required information. Therefore, these regions provide an uncertainty on which progenitors may be possible for stable mass transfer for \revtwo{C}ase A and \revtwo{C}ase B mass transfer. We already mentioned this briefly in the results, but it is important recall this when interpreting the results.

%All in all there are some uncertainties in our results as is not abnormal for research. While we do not expect these uncertainties to drastically change our conclusions it is important to keep in mind that it is possible that they do have an influence on them.

\section{Conclusion}\label{conclusion}
The goal of this research was to study the type of mass transfer that WR+O binaries might have experienced in the past and determine the mass-transfer efficiency. To do this, we used the knowledge that their progenitors are O+MS binaries and that these binaries are most likely to have a low mass ratio $q$. Based on this and the other possible restrictions discussed in \ref{Method}, we studied 21 WR+O binaries and their possible cases of mass transfer, the efficiency $\beta$, and the angular momentum loss.

To determine which case of mass transfer the WR+O binaries might have experienced, we distinguished between three different scenarios: most likely \revtwo{C}ase A, both \revtwo{C}ase A and \revtwo{C}ase B plausible, and \revtwo{C}ase A not plausible. The first scenario was the largest group in our sample. Fourteen out of \rev{21} systems most likely experienced \revtwo{C}ase A mass transfer based on our results. The other two groups contain far fewer systems.

\rev{For the systems experiencing \revtwo{C}ase A (i.e. the majority of cases), we find} the mass-transfer efficiency to have been low and no likely fixed value \citep[as found by][]{2007A&A...467.1181D}. For \revtwo{C}ase B, values up until one are possible for \rev{some} systems, while for \revtwo{C}ase A, the highest possible value we found to be possible is $\beta_{\mathrm{max}} = 0.84$. 

This is not what we expected from \reveditor{analysing} the distribution of O+MS binaries, as we only expected that about one-third of these systems undergo \revtwo{C}ase A mass transfer. Our sample may not be complete enough to include progenitors that are representative of O+MS binaries, but it may suggest that the products of \revtwo{C}ase B mass transfer do not contribute in the same way to the observed WR+O star population as post-case-A systems. \rev{Post-case-B binaries tend to have longer periods, which are harder to measure spectroscopically. However, some objects in our sample have long periods, so that the selection effect can be subtle.} One effect \rev{that might hide post- \revtwo{C}ase-B systems} may be that the stripped stars remain cool for longer \reveditor{\citep{Dutta_2024}}. Based on our results, we cannot conclude which of these phenomena is the cause of this discrepancy or whether it is caused by \rev{selection effects}.

With these results, we conclude that the observed WR+O binaries are most likely to experience \revtwo{C}ase A mass transfer, but \revtwo{C}ase B is also a possibility for some. This is not representative of which case of mass transfer we expect their progenitors (O+MS binaries) to undergo. 

The values of $\gamma$ cannot be determined accurately, but values above one are implied for most cases. However, a lower metallicity would lower the estimates of $\gamma$.

\begin{acknowledgements}
\rev{We thank the referee for detailed comments that improved the paper.} We would like to thank Alina Istrate for her help and advice on using MESA and for her advice in general. We also would like to thank Jakub Klencki for helpfully answering some of our questions. G.N. is supported by the Dutch science foundation NWO.
\end{acknowledgements}

\bibliographystyle{aa} % style aa.bst
\bibliography{sources.bib} % your references Yourfile.bib

\appendix

\section{Additional system figures}\label{appendix}
In this section additional figures for the systems not discussed in section \ref{cases-sec} can be found.

\begin{figure}
    \centering
    \includegraphics[width=0.45\textwidth]{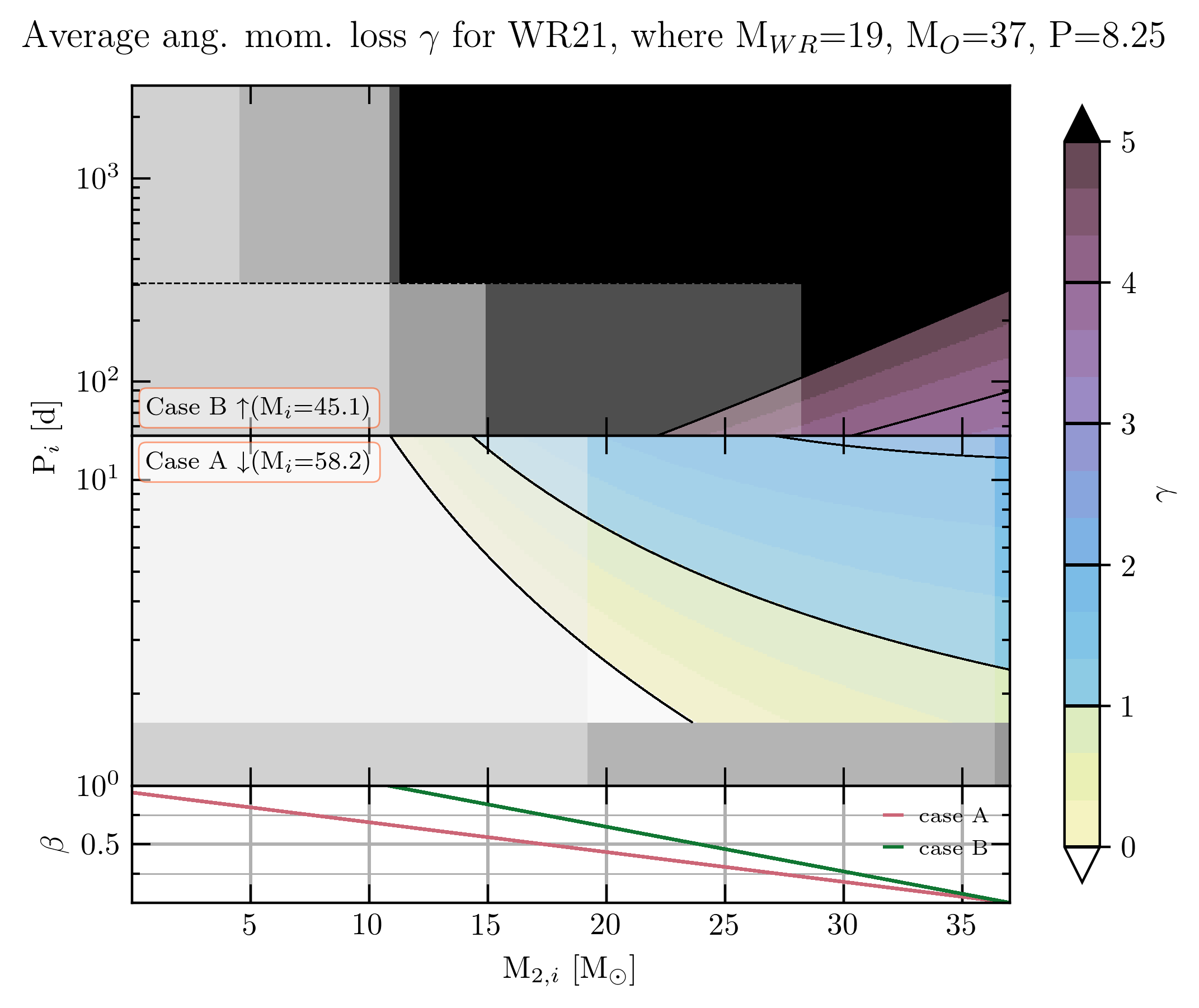}
    \vspace*{-4mm}
    \caption{The same as figure \ref{contour_WR9} but for WR21.}
    \label{contour_WR21}
\end{figure}
\vspace*{-8mm}
\begin{figure}
    \centering
    \includegraphics[width=0.45\textwidth]{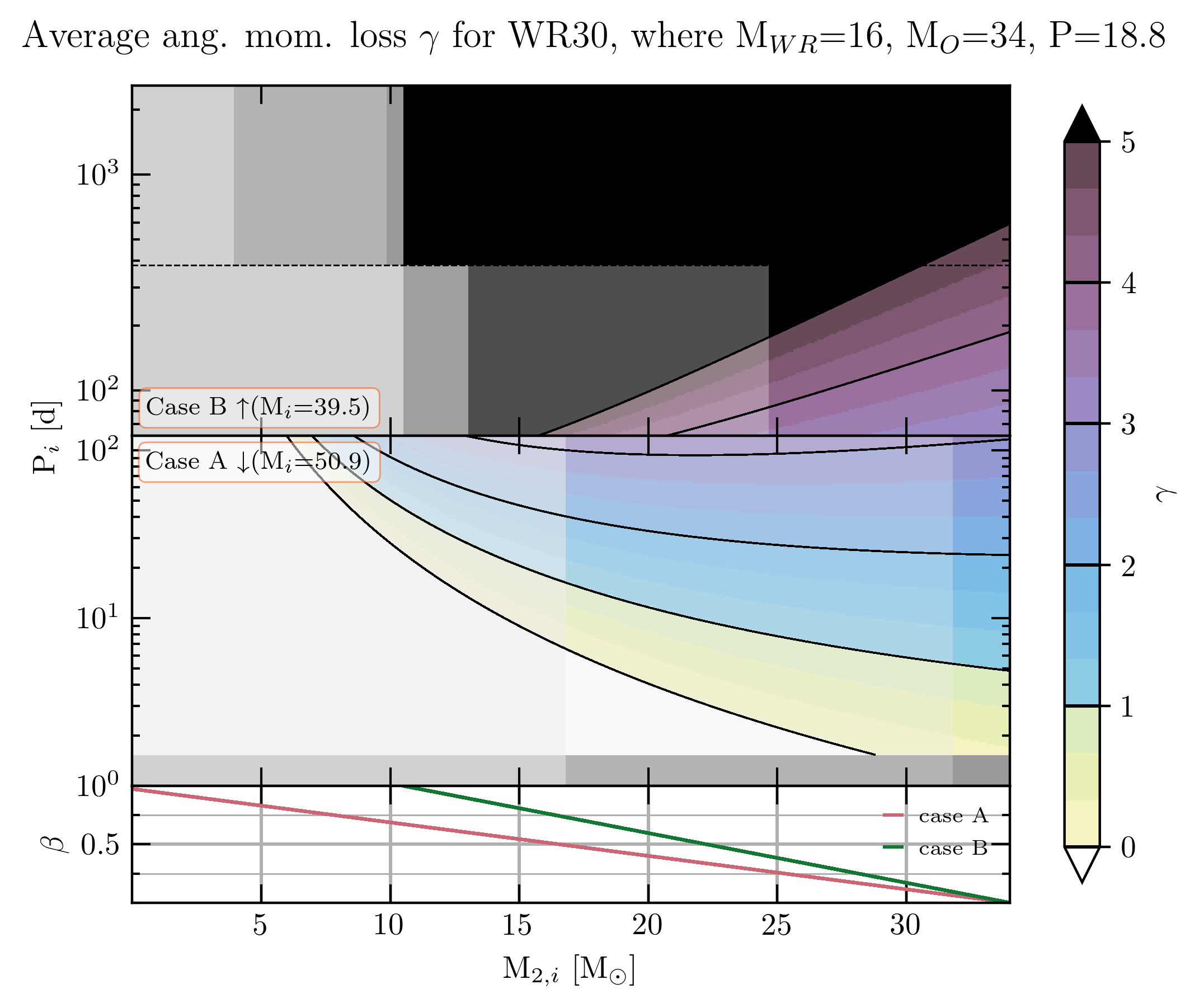}
    \vspace*{-4mm}
    \caption{The same as figure \ref{contour_WR9} but for WR30.}
    \label{contour_WR30}
\end{figure}
\vspace*{-8mm}
\begin{figure}
    \centering
    \includegraphics[width=0.45\textwidth]{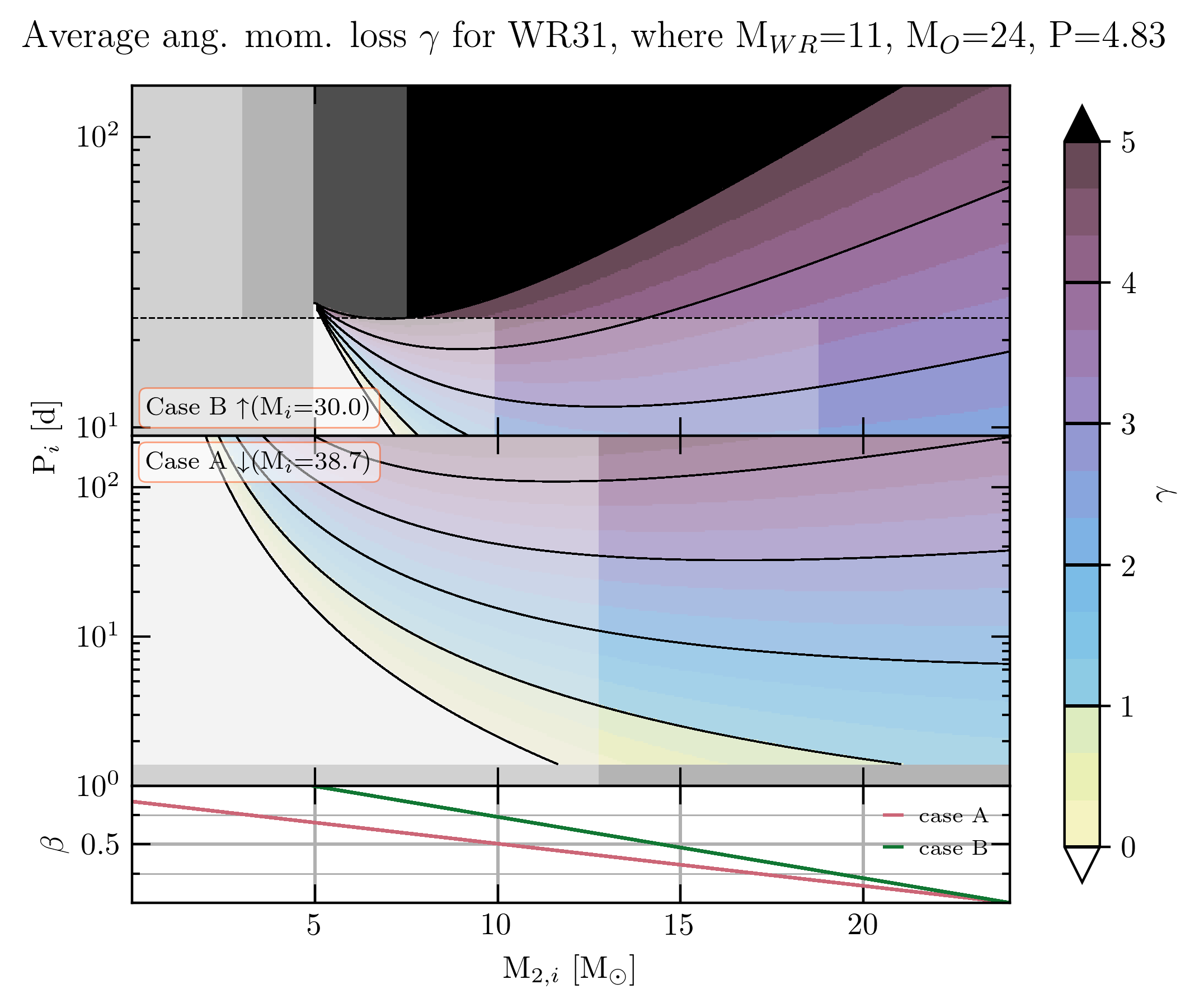}
    \vspace*{-4mm}
    \caption{The same as figure \ref{contour_WR9} but for WR31.}
    \label{contour_WR31}
\end{figure}
\vspace*{-8mm}
\begin{figure}
    \centering
    \includegraphics[width=0.45\textwidth]{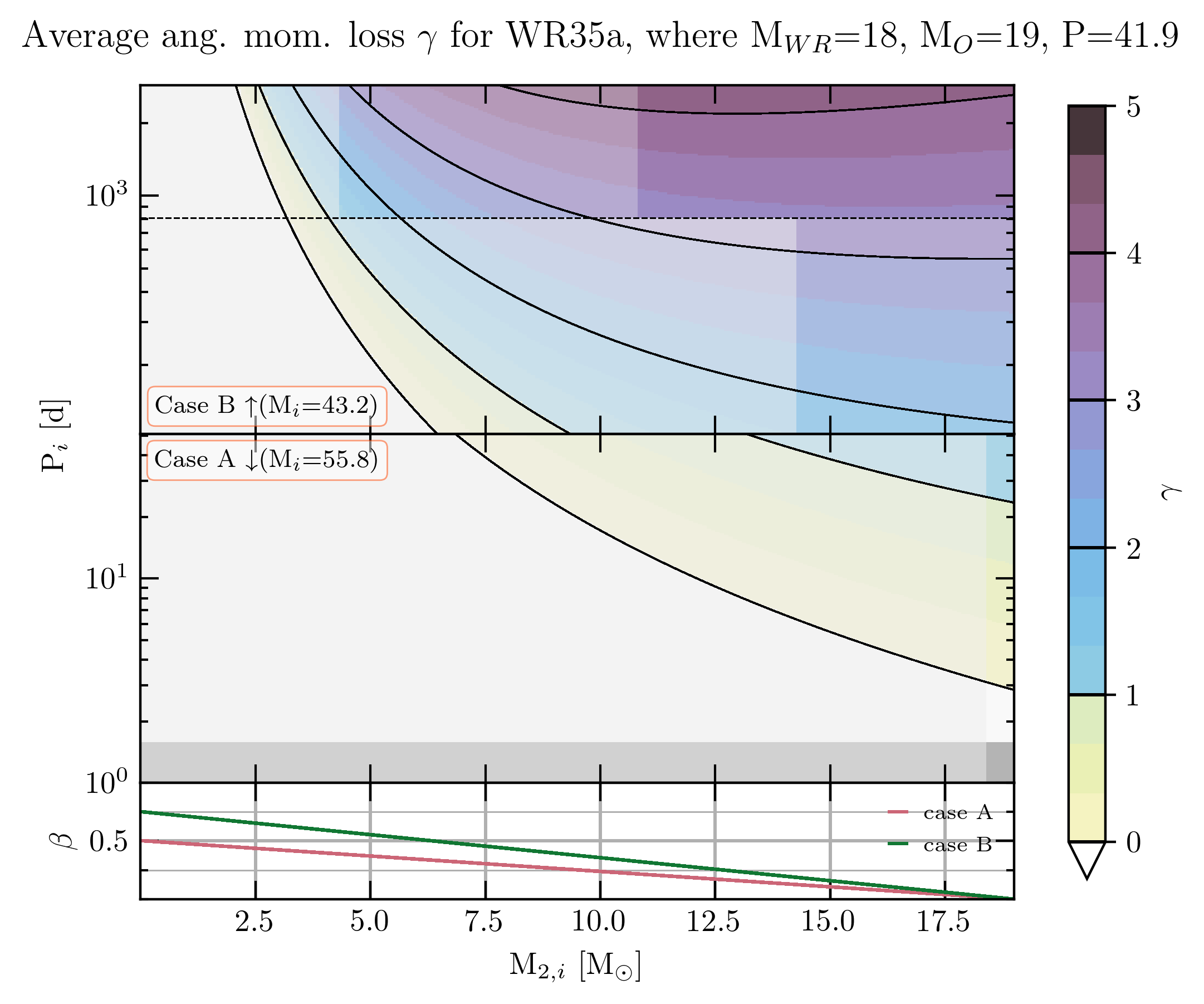}
    \vspace*{-4mm}
    \caption{The same as figure \ref{contour_WR9} but for WR35a.}
    \label{contour_WR35a}
\end{figure}
\vspace*{-8mm}
\begin{figure}
    \centering
    \includegraphics[width=0.45\textwidth]{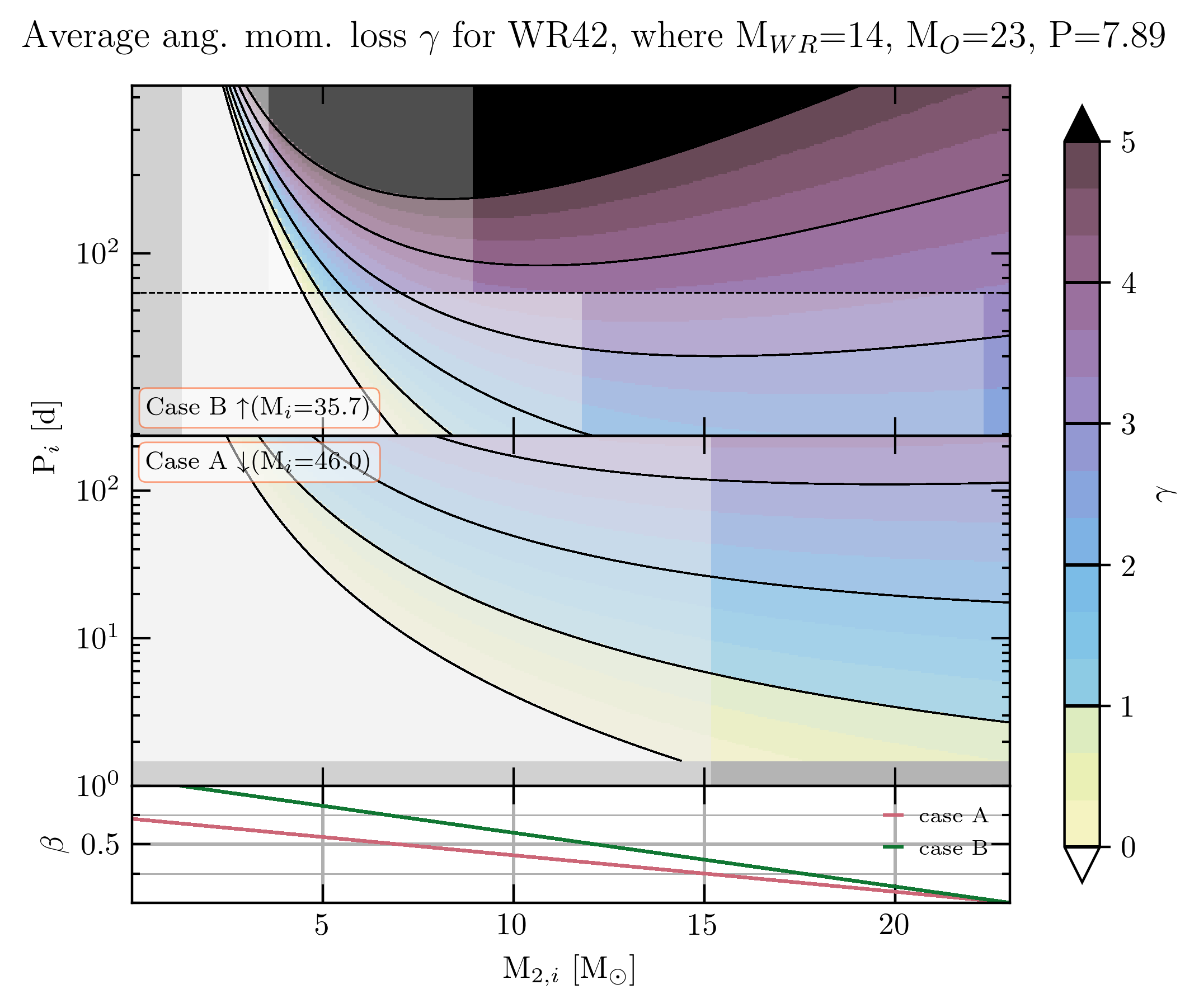}
    \vspace*{-4mm}
    \caption{The same as figure \ref{contour_WR9} but for WR42.}
    \label{contour_WR42}
\end{figure}
\vspace*{-8mm}
\begin{figure}
    \centering
    \includegraphics[width=0.45\textwidth]{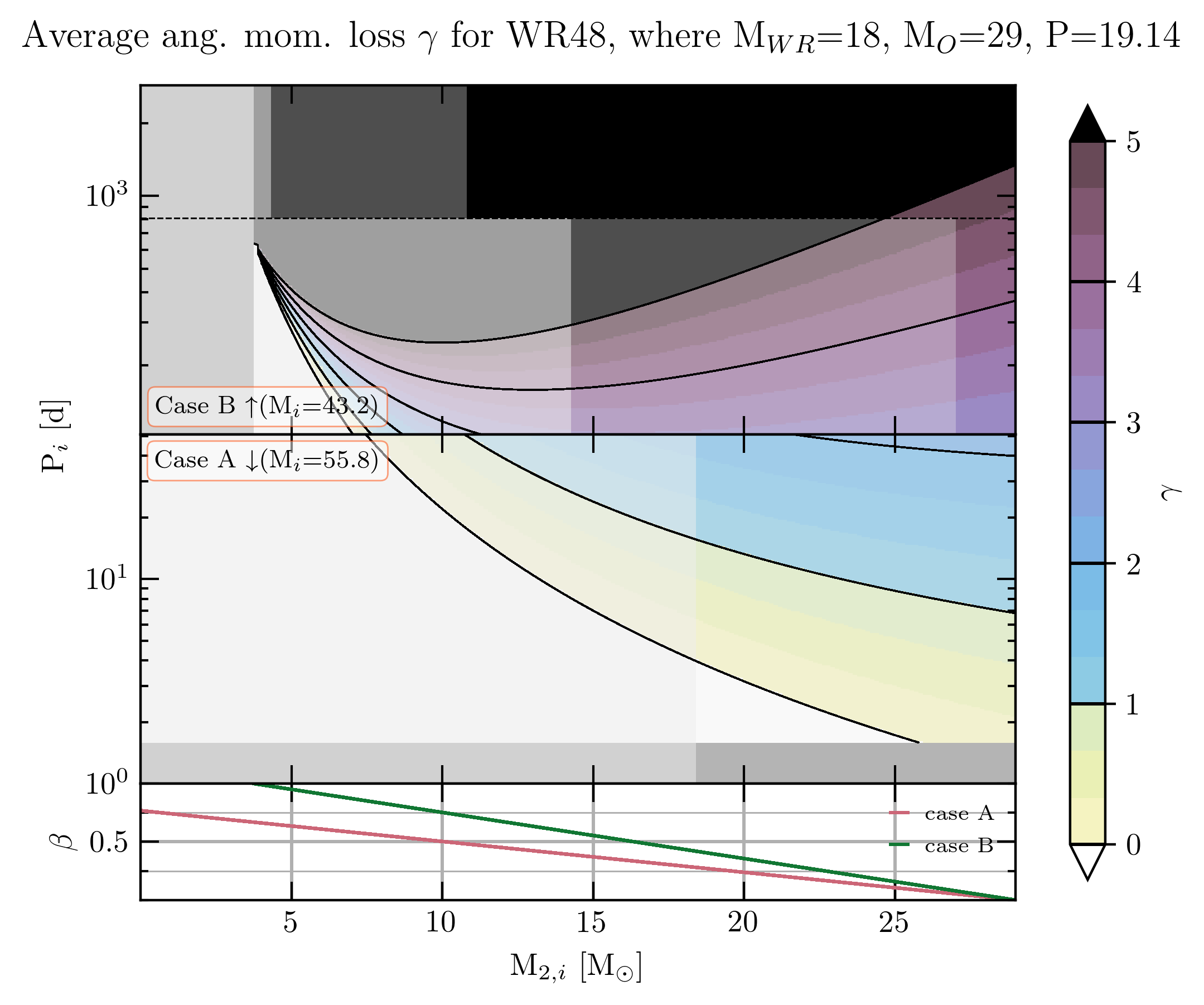}
    \vspace*{-4mm}
    \caption{The same as figure \ref{contour_WR9} but for WR48.}
    \label{contour_WR48}
\end{figure}
\vspace*{-8mm}
\begin{figure}
    \centering
    \includegraphics[width=0.45\textwidth]{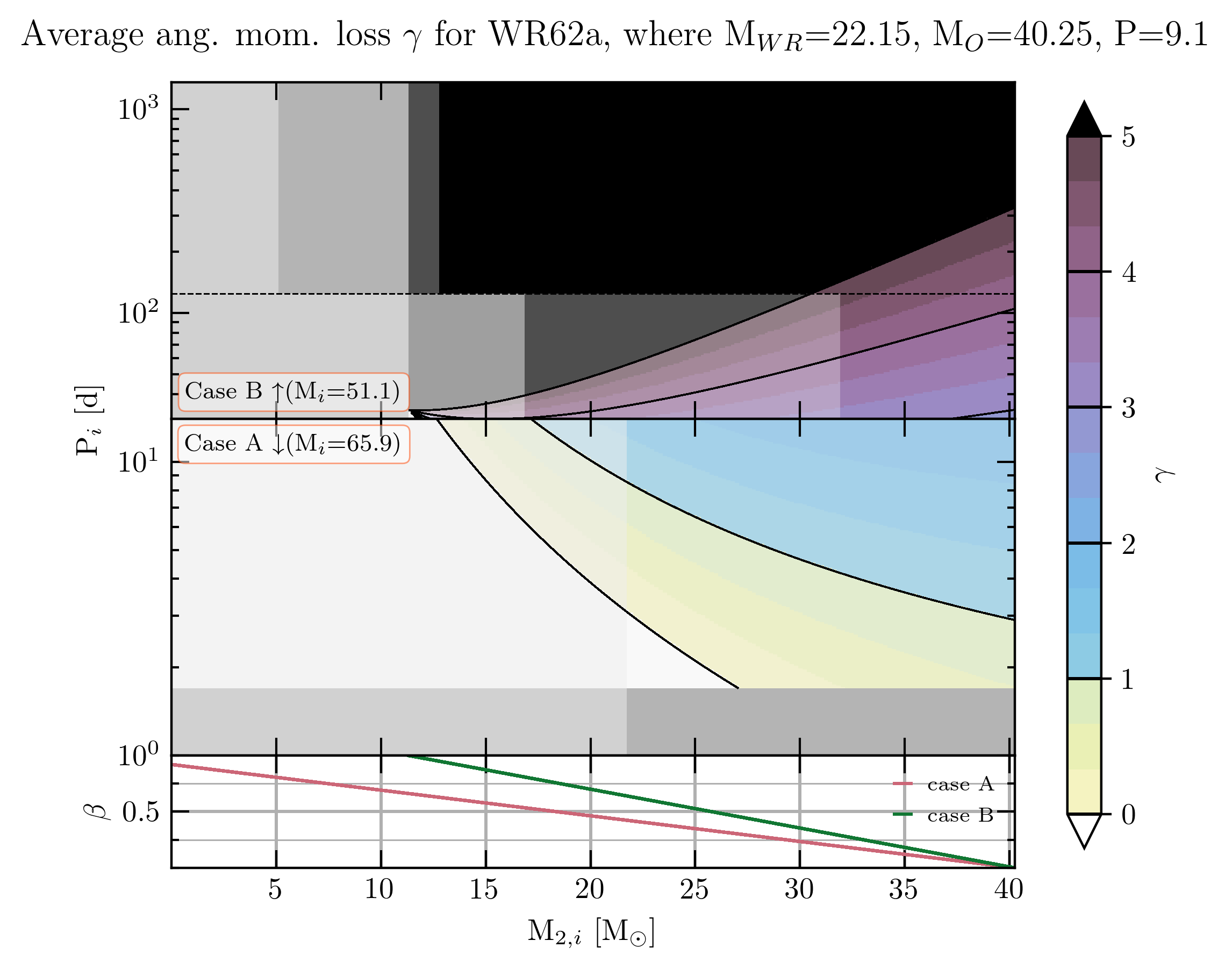}
    \vspace*{-4mm}
    \caption{The same as figure \ref{contour_WR9} but for WR62a.}
    \label{contour_WR62a}
\end{figure}
\vspace*{-8mm}
\begin{figure}
    \centering
    \includegraphics[width=0.45\textwidth]{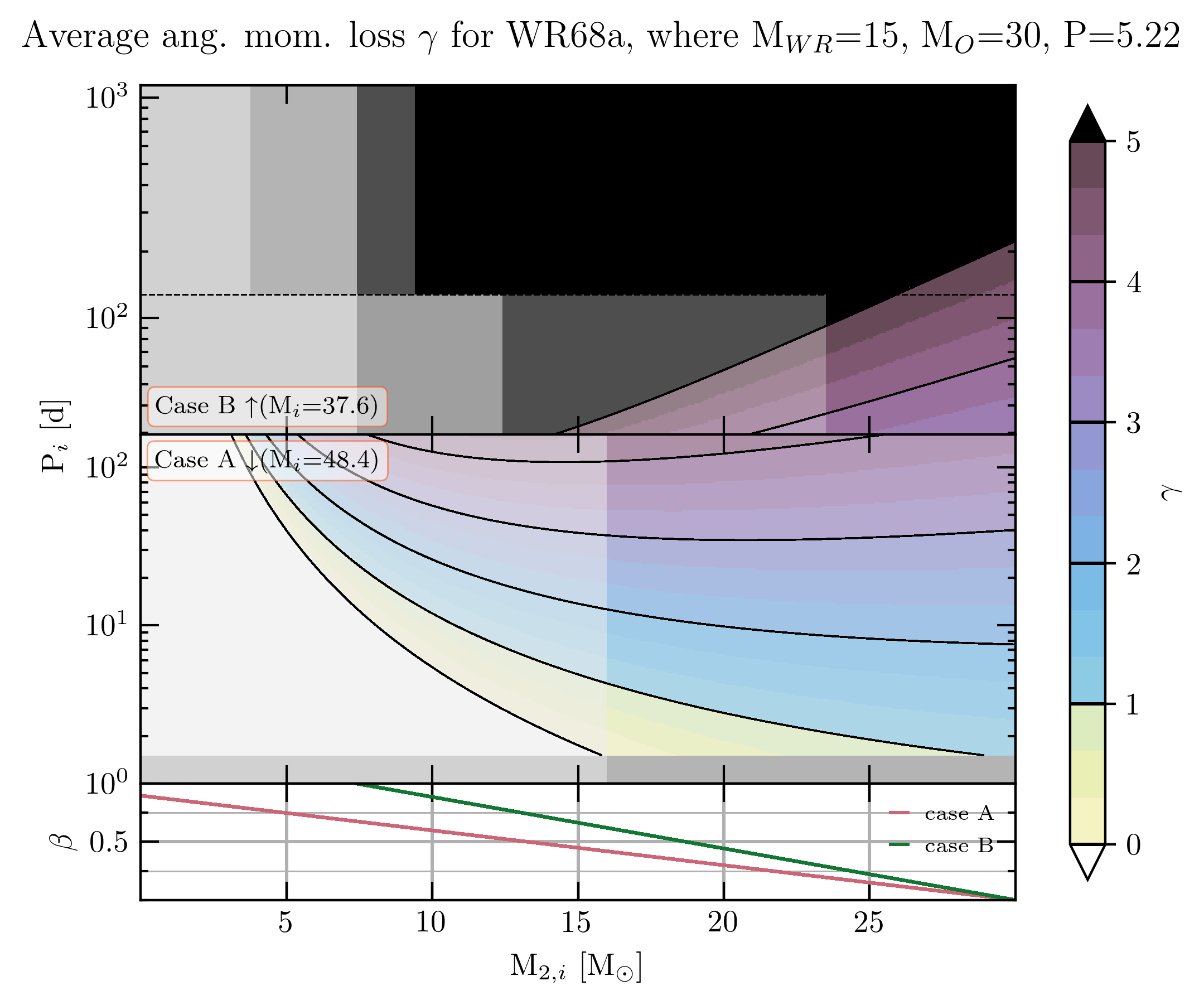}
    \vspace*{-4mm}
    \caption{The same as figure \ref{contour_WR9} but for WR68a.}
    \label{contour_WR68a}
\end{figure}
\vspace*{-8mm}
\begin{figure}
    \centering
    \includegraphics[width=0.45\textwidth]{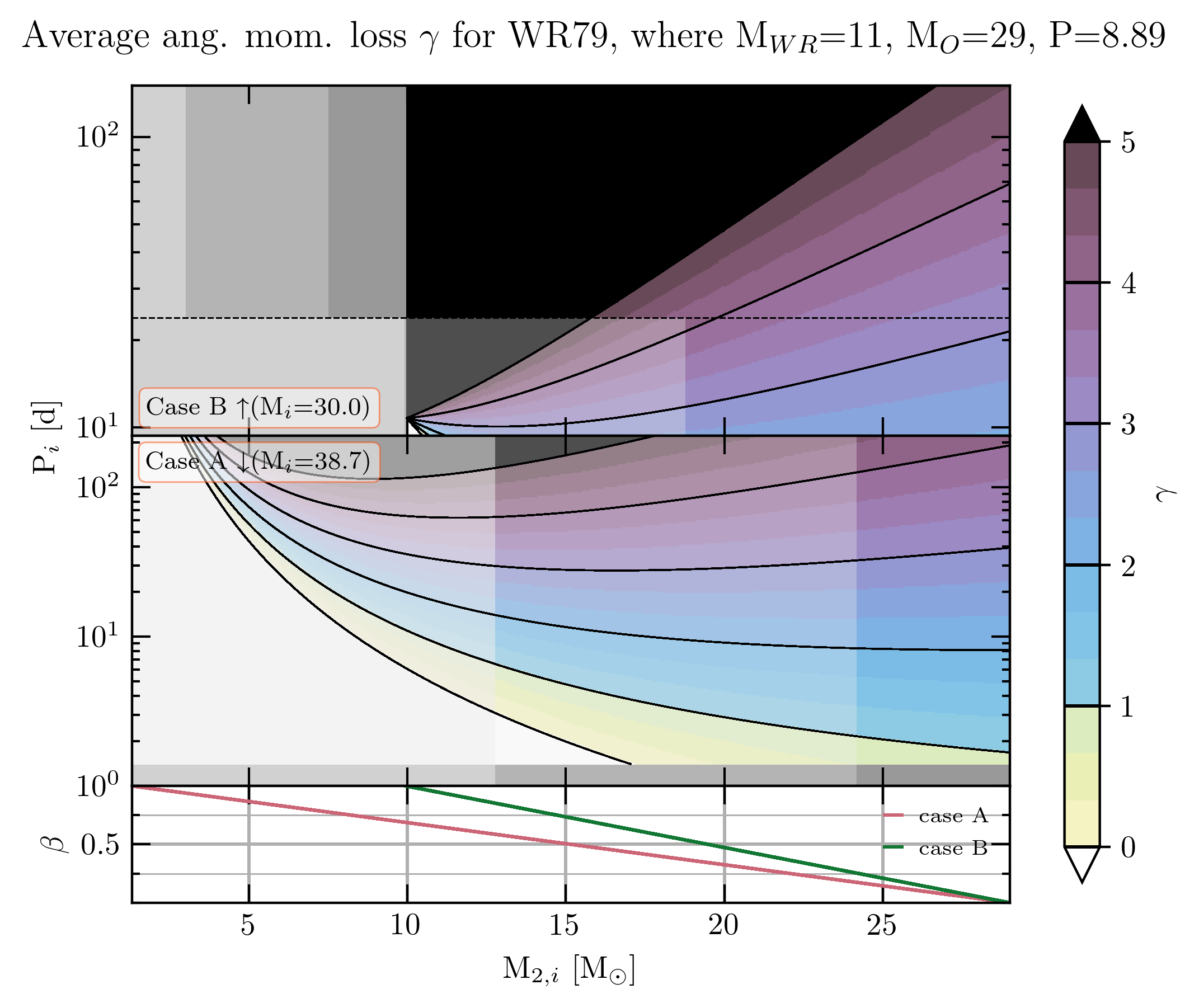}
    \vspace*{-4mm}
    \caption{The same as figure \ref{contour_WR9} but for WR79.}
    \label{contour_WR79}
\end{figure}
\vspace*{-8mm}
\begin{figure}
    \centering
    \includegraphics[width=0.45\textwidth]{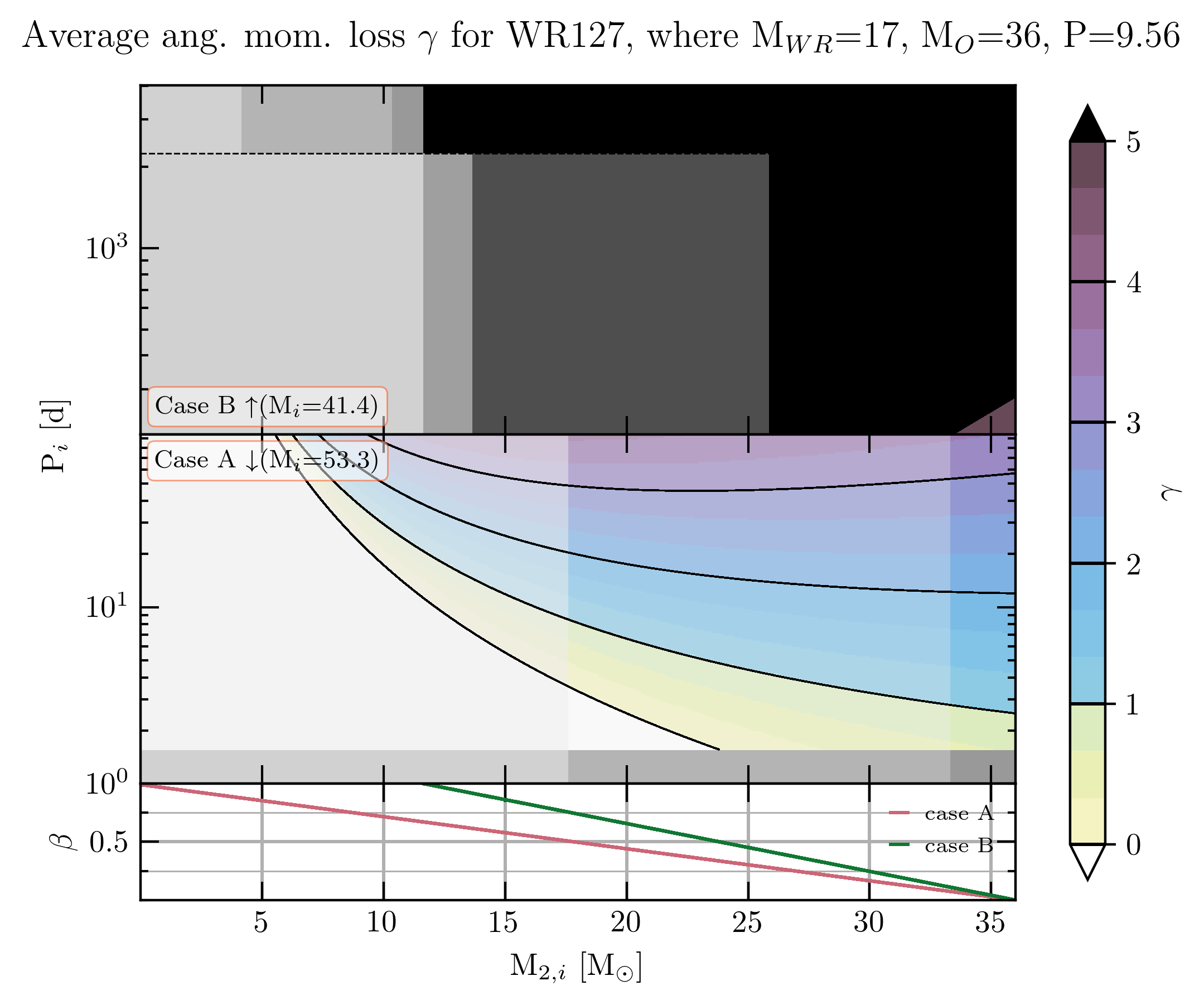}
    \vspace*{-4mm}
    \caption{The same as figure \ref{contour_WR9} but for WR127.}
    \label{contour_WR127}
\end{figure}
\vspace*{-8mm}
\begin{figure}
    \centering
    \includegraphics[width=0.45\textwidth]{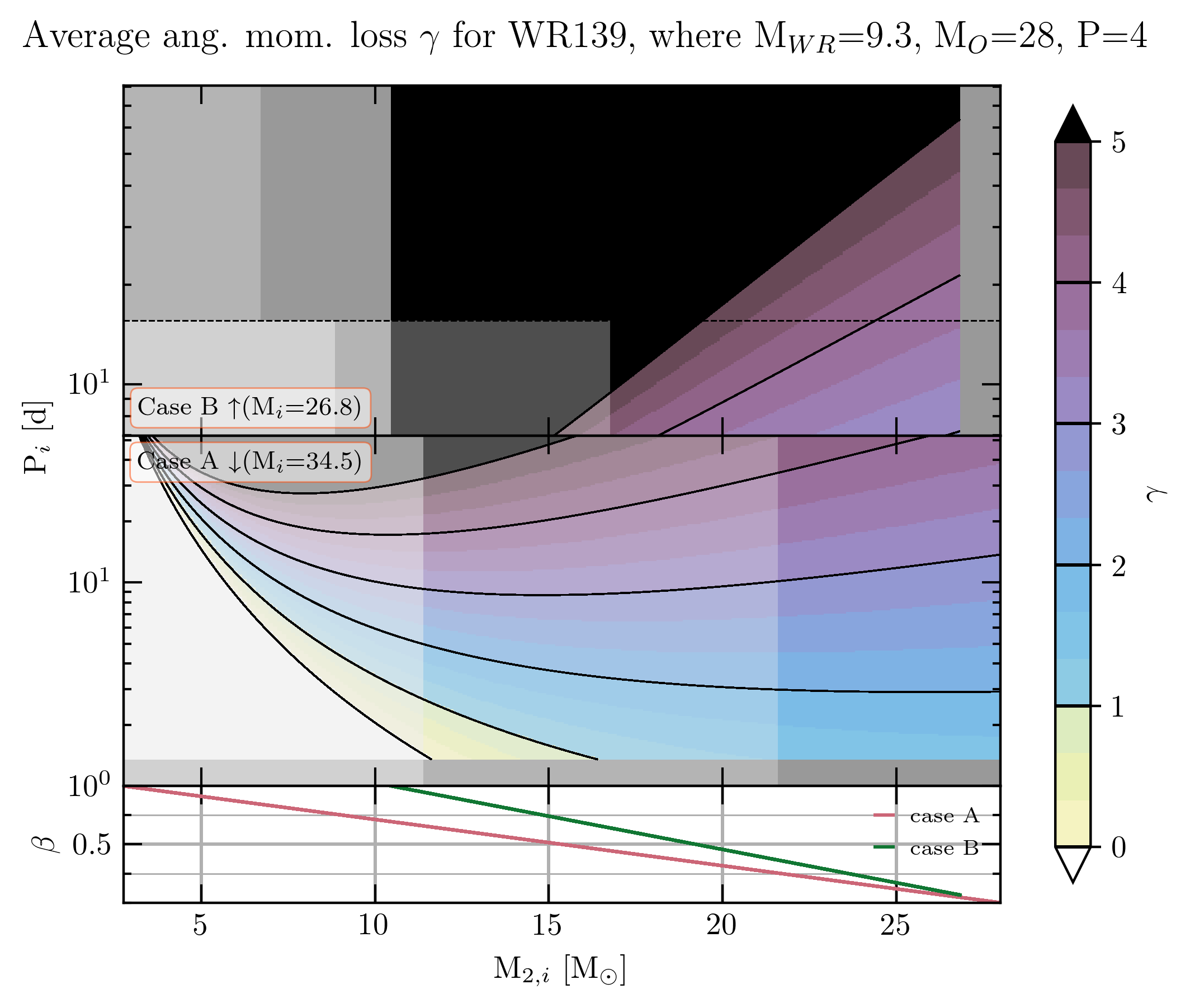}
    \vspace*{-4mm}
    \caption{The same as figure \ref{contour_WR9} but for WR139.}
    \label{contour_WR139}
\end{figure}
\vspace*{-8mm}
\begin{figure}
    \centering
    \includegraphics[width=0.45\textwidth]{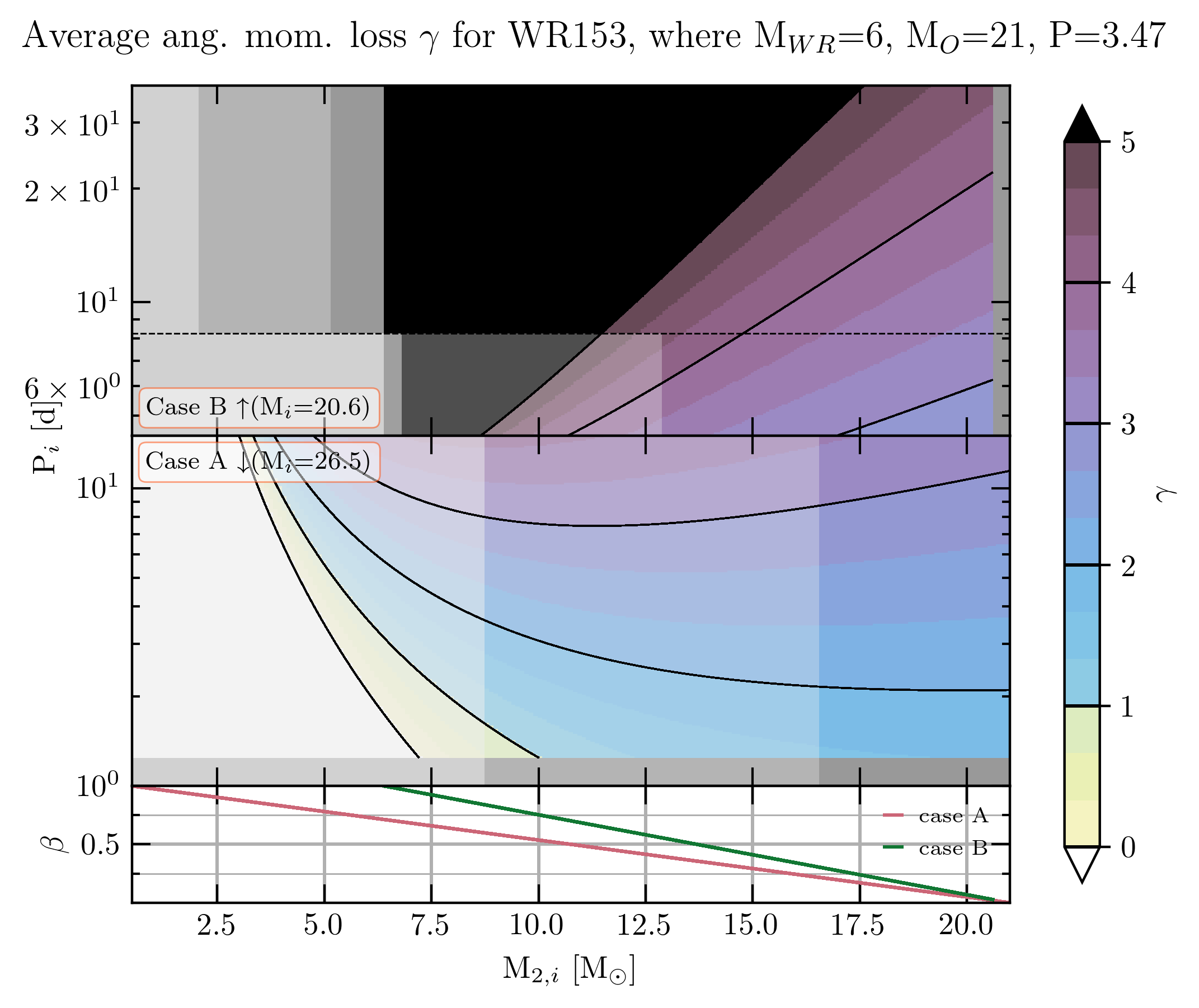}
    \vspace*{-4mm}
    \caption{The same as figure \ref{contour_WR9} but for WR153.}
    \label{contour_WR153}
\end{figure}

\begin{figure}
    \centering
    \includegraphics[width=0.45\textwidth]{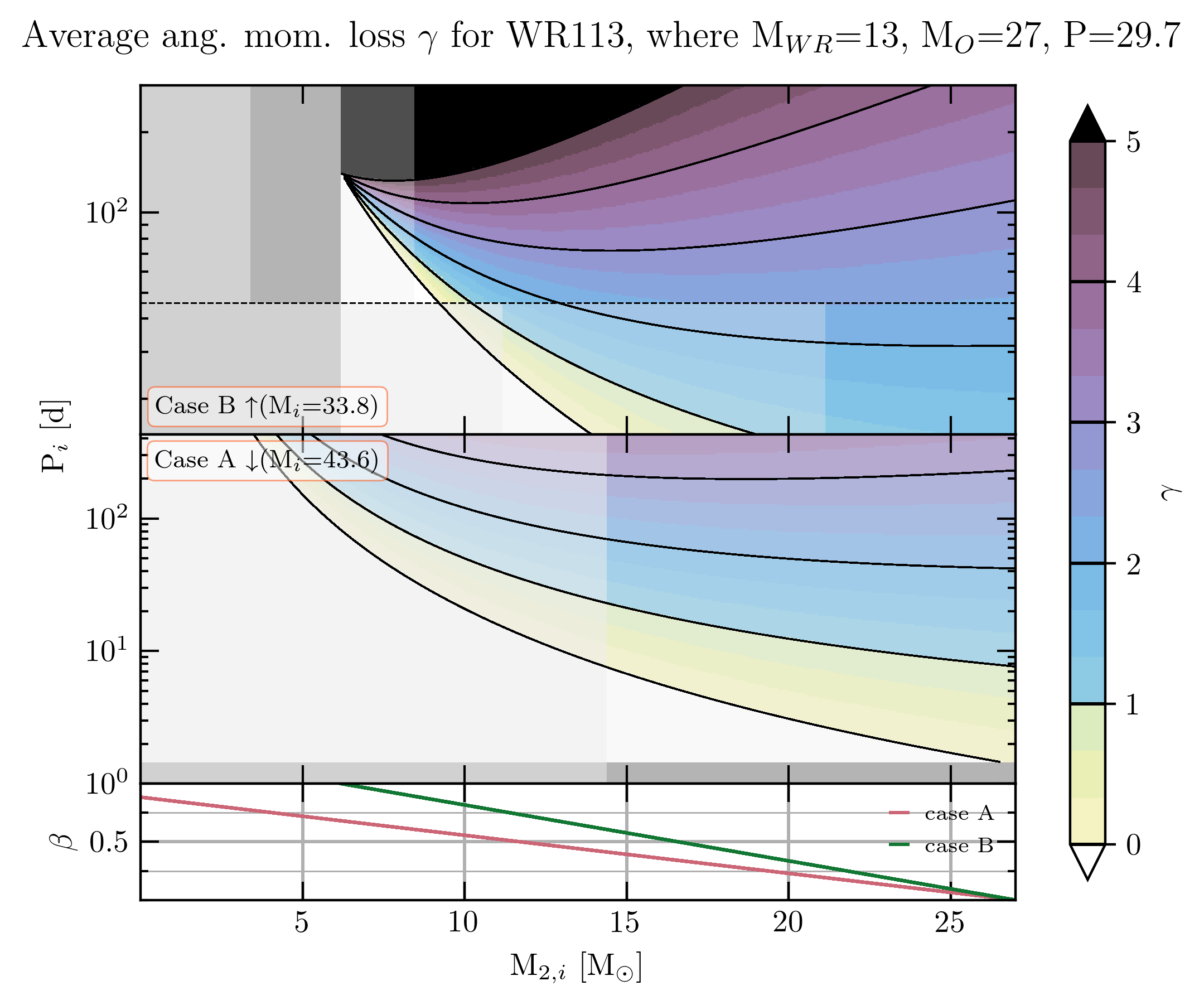}
    \vspace*{-4mm} 
    \caption{The same as figure \ref{contour_WR9} but for WR113.}
    \label{contour_WR113}
\end{figure}
\vspace*{-8mm}
\begin{figure}
    \centering
    \includegraphics[width=0.45\textwidth]{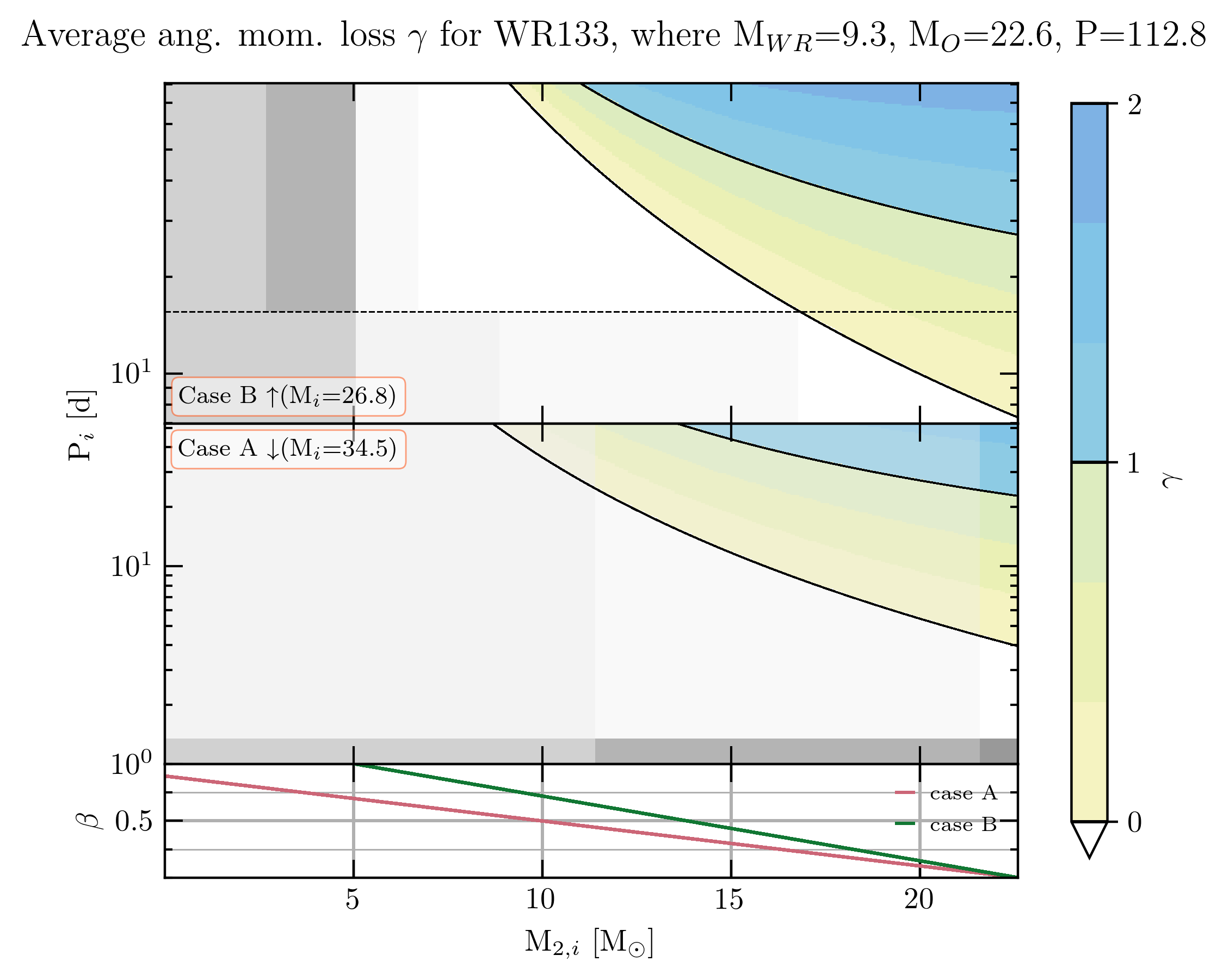}
    \vspace*{-4mm}
    \caption{The same as figure \ref{contour_WR9} but for WR133.}
    \label{contour_WR133}
\end{figure}

\begin{figure}
    \centering
    \includegraphics[width=0.45\textwidth]{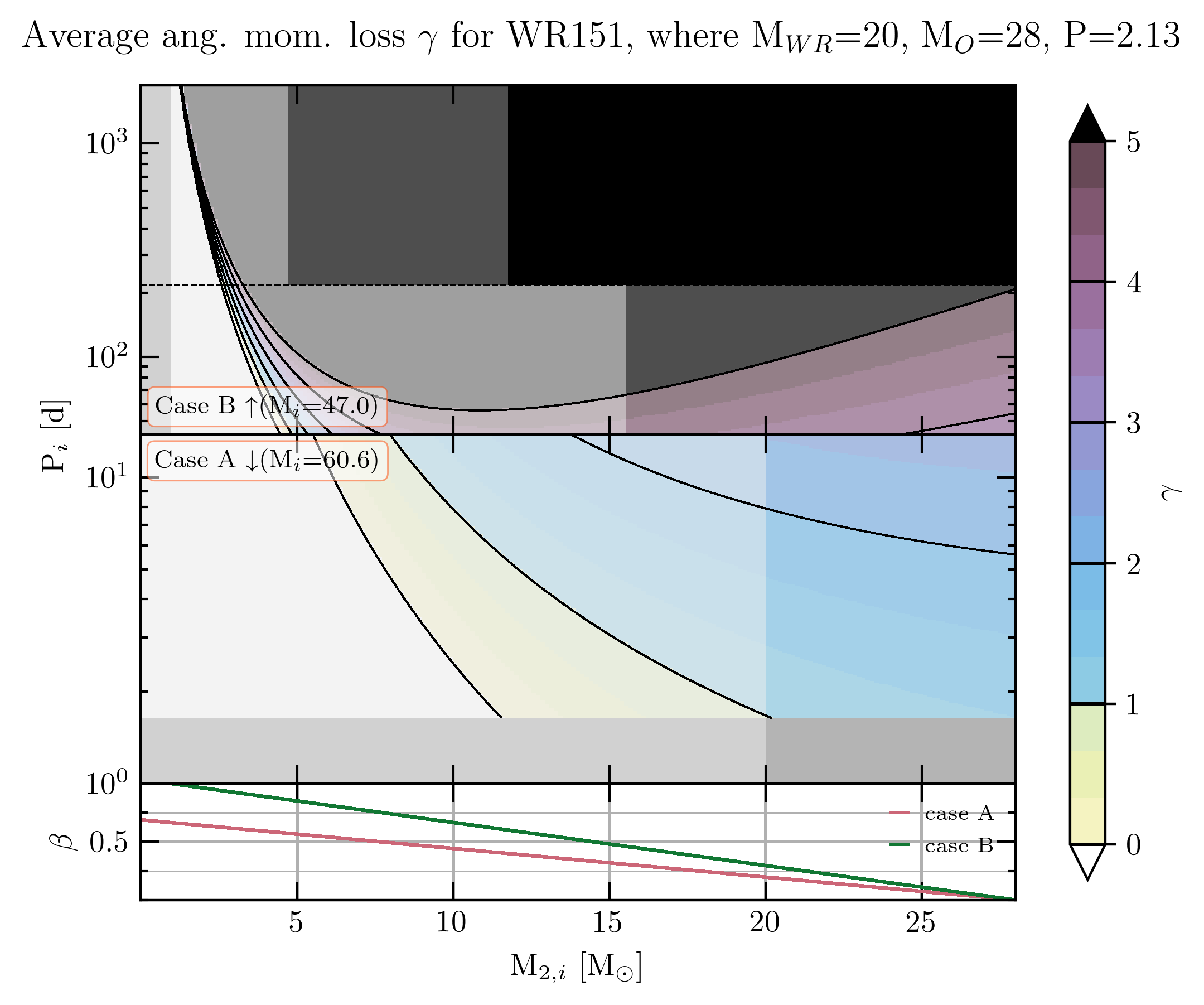}
    \vspace*{-4mm}
    \caption{The same as \ref{contour_WR9} but for WR151.}
    \label{contour_WR151}
\end{figure}
\vspace*{-8mm}
\begin{figure}
    \centering
    \includegraphics[width=0.45\textwidth]{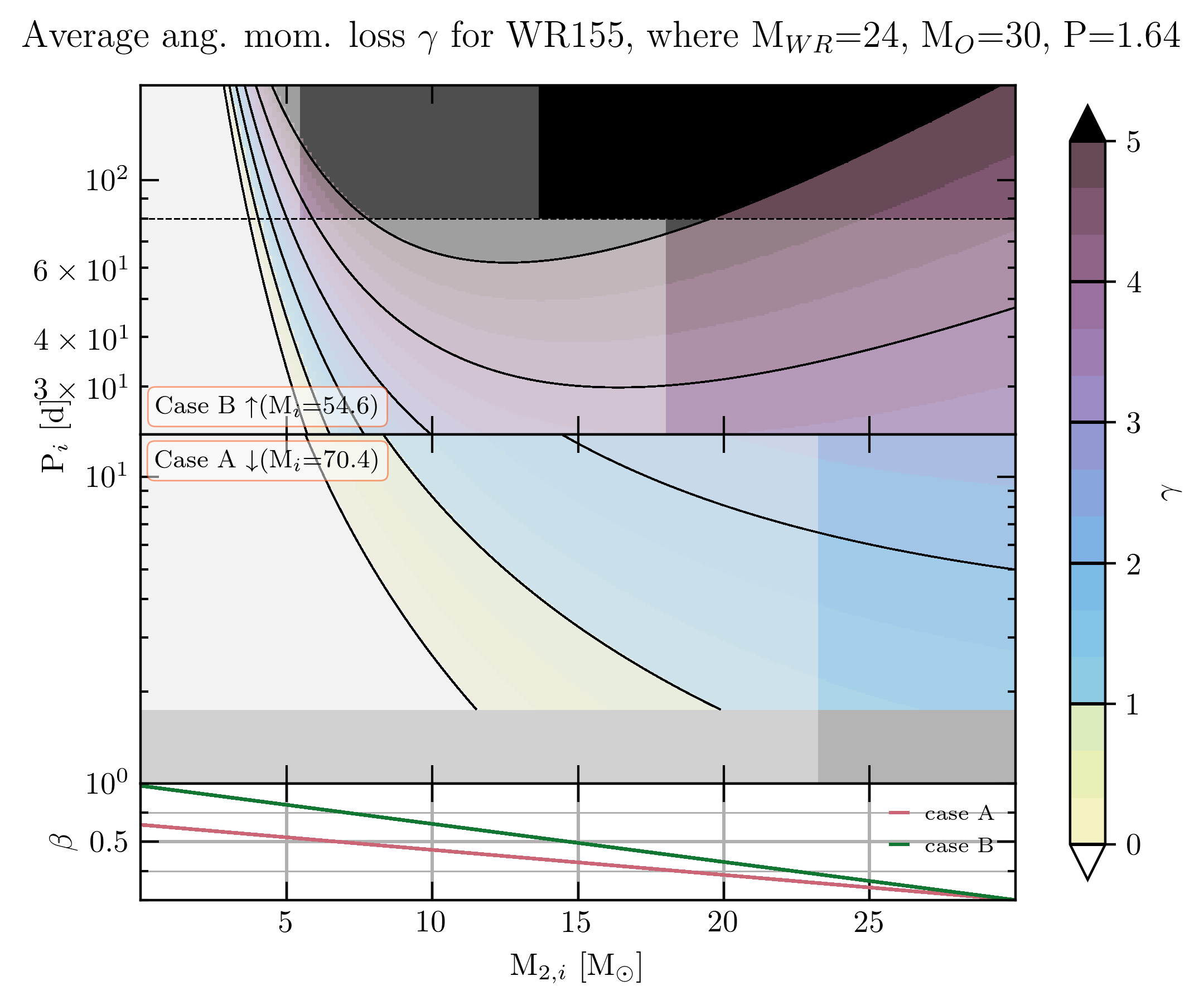}
    \vspace*{-4mm}
    \caption{The same as \ref{contour_WR9} but for WR155.}
    \label{contour_WR155}
\end{figure}

\end{document}